\documentclass[11pt]{article}

\usepackage[margin=1in]{geometry}
\usepackage{amsmath,amssymb,amsthm}
\usepackage{bm}
\usepackage{graphicx}
\usepackage{float}
\usepackage{subcaption}
\usepackage{booktabs}
\usepackage{adjustbox}
\usepackage{hyperref}
\usepackage{microtype}
\usepackage{enumitem}
\usepackage{mathtools}
\usepackage{natbib}
\newcommand{\R}{\mathbb{R}}
\newcommand{\E}{\mathbb{E}}
\newcommand{\Var}{\operatorname{Var}}
\newcommand{\Cov}{\operatorname{Cov}}
\newcommand{\Corr}{\operatorname{Corr}}
\newcommand{\1}{\mathbf{1}}

\title{Beyond Picking Winners: Correlation-Driven Tail Risk in Venture Capital Portfolio Construction}
\author{
  Yunqi Liang\textsuperscript{1} \and
  Hasan Ugur Koyluoglu\textsuperscript{2} \and
  Fuat Alican\textsuperscript{3} \and
  Yigit Ihlamur\textsuperscript{3}
}

\date{
  \textsuperscript{1}University of Oxford \quad
  \textsuperscript{2}Oliver Wyman \quad
  \textsuperscript{3}Vela Research \\
}

\begin{document}
\maketitle

\begin{abstract}
We propose a Gaussian-copula-based framework that learns deal-level dependence directly from observed joint success frequencies across founder, geography, and market attributes. Holding marginal deal success probabilities fixed, deal-level correlation preserves expected portfolio outcomes but shifts the portfolio distribution toward heavier right tails and higher kurtosis. In portfolio simulations, correlation reduces the probability of modest success counts while sharply amplifying extreme upside outcomes, especially in structurally concentrated portfolios. Our findings suggest that extreme venture capital outcomes may partly reflect correlation-induced tail amplification rather than solely higher average deal quality, with potential implications for portfolio construction and risk management. We note that the observed dataset reflects selected deals with observable outcomes, which inflates apparent success rates relative to the true population base rate; however, the core finding that correlation reshapes the distributional shape while leaving the mean unchanged is structurally robust to the level of marginal success probabilities.
\end{abstract}

\section{Introduction}
Venture capital (VC) returns are well-known to be heavy-tailed, with a small number of extreme outcomes driving a large portion of aggregate performance. A central question in portfolio construction is whether this heavy-tailedness arises mainly from heterogeneity in marginal deal quality or from dependence across deals. We develop a unified Gaussian copula model that learns deal-level dependence from observed joint success frequencies across three attribute families: founder type, geography, and market category. The model is flexible enough to capture dependencies between groups such as founder-market or geography-market interactions. It also enables scalable estimation through moment matching of joint probabilities.

Empirically, we find that latent deal correlation has limited impact on the expected number of successful deals, but it substantially increases dispersion, skewness, and kurtosis. Portfolio simulations further show that correlation redistributes probability mass, by reducing the likelihood of achieving a small number of successes while amplifying the extreme upper tail. This mechanism explains why concentrated portfolios can exhibit much larger upside, at the cost of higher tail risk.

An important calibration caveat is addressed directly in Section 4.2. Our dataset reflects deals with observable outcomes, which introduces selection bias that inflates apparent success rates to approximately 9\%, well above the population-level base rate of roughly 1.9\% to 2\% documented in the broader VC literature. The key contribution of this paper concerns the relative effect of correlation on portfolio distributions, specifically how dependence shifts variance, skewness, and kurtosis while leaving the mean unchanged. This relative comparison is structurally robust to the level of marginal success probabilities, because the Gaussian copula framework separates marginal rates from the dependence structure. Holding the correlation structure ($\Sigma$) fixed and scaling marginal probabilities toward population-level rates would compress absolute success counts but preserve the qualitative finding that correlation amplifies tail probabilities relative to independence.

The remainder of the paper is organized as follows. Section 2 reviews the related literature. Section 3 details the methodology for constructing and estimating deal-level correlations. Section 4 presents the empirical analysis, including the construction of synthetic deal probabilities, the estimation of the dependence structure, and the resulting correlation distributions. Section 5 reports the portfolio simulation results for different portfolio types, with a focus on the distribution of successful outcomes. Section 6 concludes.

\section{Literature Review}

VC investment outcomes are widely documented to be highly skewed and heavy-tailed, with a small number of outsized successes driving aggregate performance. \citet{cochrane2005venture} provides a foundational selection-corrected analysis of VC returns and shows that high volatility and strong positive skewness persist even after accounting for selective valuation reporting. Beyond heavy tails, VC payoffs also exhibit meaningful cross-sectional dependence, complicating both performance evaluation and portfolio construction. Using stochastic discount factor methods, \citet{korteweg2016risk} highlight several distinctive features of venture returns, including infrequent realizations, skewness, endogenous horizons, and cross-sectional dependence.

A related literature examines how diversification shapes risk-taking and return distributions in VC portfolios. \citet{buchner2017diversification} show that diversification affects downside and upside risk asymmetrically and is linked to endogenous selection into riskier investments, reinforcing the importance of non-normality and skewness in venture outcomes. More recent evidence emphasizes portfolio composition. \citet{teti2024diversification} find that larger portfolios with limited industry diversification can still perform well, while diversification across stages and geographic regions appears less consequential.

Our dependence modeling builds on copula theory, which separates marginal event probabilities from the dependence structure governing joint realizations \citep{sklar1959fonctions, nelsen2006copulas}. In portfolio-event settings, copula and latent-factor approaches have been widely developed in credit risk modeling. \citet{koyluoglu1998reconcilable} study default correlation in portfolio credit risk, \citet{li2000copula} introduces a copula-based framework for modeling joint defaults, and \citet{gordy2003risk} shows that common-factor structures can generate portfolio-invariant risk contributions. \citet{frey2003dependent} further emphasize latent-variable copula models. For high-dimensional settings, \citet{oh2017factorcopula} propose scalable factor-copula models and document stronger co-movement during extreme market states.

Building on these literatures, we model the dependence between VC deals using a Gaussian copula framework in which outcomes are driven by heterogeneous marginal success probabilities and a learned dependence structure. Our framework provides a tractable first-pass approach to studying how deal-level dependence contributes to the skewness and kurtosis observed in VC portfolio returns. We note that the Gaussian copula has an upper tail dependence coefficient of exactly zero for any finite latent correlation, which is a known limitation in heavy-tailed environments. Extending to copulas with non-zero upper tail dependence, such as the t-copula or Gumbel family, is a natural direction for future work and would more directly capture extreme joint success events of the type that define top-decile VC fund performance.

\section{Methodology}
To capture the dependence structure across deals, we adopt a unified Gaussian-copula framework. The section details the latent Gaussian structure, implied dependence, and estimation procedure.
\subsection{Setup and notation}
Consider a collection of deals indexed by $i=1,\dots,N$. Each deal is associated with three types of categorical attributes:
\begin{itemize}[leftmargin=3.6em]
  \item \textbf{Founder type}: 2 mutually exclusive categories;
  \item \textbf{Geography}: 4 mutually exclusive categories;
  \item \textbf{Market category}: 6 categories; multiple may apply to a single deal.
\end{itemize}

We encode attributes as a single binary indicator vector
\[
e_i \in \{0,1\}^{d}, \qquad d = 2 + 4 + 6 = 12,
\]
constructed as
\[
e_i =
\begin{bmatrix}
F_i \\
G_i \\
M_i
\end{bmatrix},
\]
where $F_i$ is a 2-dimensional one-hot vector, $G_i$ is a 4-dimensional one-hot vector, and $M_i$ is a 6-dimensional multi-hot vector. This encoding allows each deal to load on multiple market dimensions while remaining exclusive in founder type and geography.

\paragraph{Remark on encoding constraints.} Because founder type and geography are mutually exclusive, the corresponding sub-vectors satisfy sum-to-one constraints: $F_{1, i} + F_{2, i} = 1$ and $G_{1, i} + G_{2, i} + G_{3, i} + G_{4, i} = 1$ for every deal $i$. This introduces linear dependencies within each one-hot block, implying that the 12-dimensional attribute vector does not span a full-rank space. As a consequence, $\Sigma$ is not uniquely identified in the unrestricted sense: many matrices may produce equivalent fits to the observed joint frequencies. In particular, signs and magnitudes of cross-block elements involving the first-time founder indicator should be interpreted with this constraint in mind rather than as purely economic signals. A natural refinement is to drop one reference category per exclusive group, consistent with standard dummy variable treatment in regression; we leave this for future work.

\subsection{Latent Gaussian structure}
\paragraph{Attribute-level latent factors.}

We introduce a latent Gaussian attribute vector
\[
F \sim \mathcal{N}(0,\Sigma),
\]
where $F\in\R^d$ and $\Sigma\in\R^{d\times d}$ is a positive semidefinite covariance matrix. To ensure positive semidefiniteness during estimation, we parameterize
\[
\Sigma = L L^\top,
\]
with $L\in\R^{d\times k}$, allowing an optional low-rank structure. Importantly, $\Sigma$ is fully general and not block-diagonal, permitting unrestricted second-order dependence across founder, geography, and market attributes.

\paragraph{Deal-level latent variable.}

For each deal $i$, define a scalar latent variable
\begin{equation}
Z_i = \alpha_0 U + e_i^\top F + \phi_i \varepsilon_i, \label{eq:Zi}
\end{equation}
where $U\sim\mathcal{N}(0,1)$ is a global common factor, $\varepsilon_i\sim\mathcal{N}(0,1)$ are idiosyncratic shocks, and $\alpha_0\in[0,1)$ controls global correlation. To normalize marginal variances, choose
\[
\phi_i^2 = 1-\alpha_0^2 - e_i^\top \Sigma e_i,
\]
so that $\Var(Z_i)=1$. Equivalently, writing $F=LP$ with $P\sim\mathcal{N}(0,I_k)$,
\[
Z_i = \alpha_0 U + e_i^\top L P + \sqrt{1-\alpha_0^2-e_i^\top\Sigma e_i}\,\varepsilon_i.
\]

\subsection{Covariance structure}
For any two deals $i,j$, the pairwise latent covariance is given by
\begin{equation}
\Cov(Z_i,Z_j) = \alpha_0^2 + e_i^\top \Sigma e_j. \label{eq:cov}
\end{equation}
This defines the core structural kernel of the model.

Partition $\Sigma$ according to attribute groups:
\[
\Sigma =
\begin{pmatrix}
\Sigma_{FF} & \Sigma_{FG} & \Sigma_{FM} \\
\Sigma_{GF} & \Sigma_{GG} & \Sigma_{GM} \\
\Sigma_{MF} & \Sigma_{MG} & \Sigma_{MM}
\end{pmatrix}.
\]
Then $\Sigma_{FF}$ captures founder-founder dependence, $\Sigma_{GG}$ geography-geography, $\Sigma_{MM}$ market-market, while $\Sigma_{FG}$, $\Sigma_{FM}$, $\Sigma_{GM}$ encode explicit cross-group interactions.

\subsection{Observation model (Gaussian copula)}
Each deal has a binary outcome $X_i\in\{0,1\}$. Let $p_i=\mathbb{P}(X_i=1)$ denote the marginal success probability, and define threshold $t_i=\Phi^{-1}(p_i)$ where $\Phi$ is the standard normal cumulative distribution function (CDF). Observations are generated as
\begin{equation}
X_i = \1[Z_i \le t_i]. \label{eq:obs}
\end{equation}

\subsection{Exact joint probabilities and induced Bernoulli correlation}
For any pair $(i,j)$,
\[
\mathbb{P}(X_i=1,X_j=1)=\Phi_2(t_i,t_j;r_{ij}),
\]
where $\Phi_2$ is the bivariate normal CDF and
\[
r_{ij}=\Corr(Z_i,Z_j)=\alpha_0^2 + e_i^\top \Sigma e_j \in (-1,1).
\]
The induced Bernoulli correlation is
\[
\Corr(X_i,X_j) =
\frac{\Phi_2(t_i,t_j;r_{ij}) - p_i p_j}
{\sqrt{p_i(1-p_i)\,p_j(1-p_j)}}.
\]

\subsection{Estimation from empirical joint probabilities}
Let $\mathcal{P}$ denote a set of sampled deal pairs. The empirical estimator of $\Phi_2(t_i,t_j; r_{ij}(\theta)) = \E[X_iX_j]$ can be formed from observed outcomes. For each attribute pair $(u,v)$, we aggregate over sampled deal pairs $\{(i_k,j_k)\}_{k=1}^K$, then the empirical joint probability is given by
\begin{equation}
\widehat{J}^{\text{emp}}_{uv} = \frac{1}{K}\sum_{k=1}^K x_{i_k}x_{j_k},
\label{eq:emp_joint_prob_equ}
\end{equation}
and the model joint probability is defined as 
\begin{equation}
\widehat{J}^{\text{model}}_{uv}(\theta)
= \frac{1}{K}\sum_{k=1}^K \Phi_2(t_{i_k},t_{j_k}; r_{i_k j_k}(\theta)),
\label{eq:model_joint_prob_equ}
\end{equation}
where $\theta=(\alpha_0,\Sigma)$. Direct evaluation of $\Phi_2$ is computationally expensive. For large-scale fitting, we could use the first-order approximation around $r_{ij}=0$ for the bivariate normal CDF by $
\Phi_2(t_i,t_j;r_{ij}) = \Phi(t_i)\Phi(t_j) + r_{ij}\phi(t_i)\phi(t_j) + O(r_{ij}^2),
$
where $\phi$ is the standard normal probability density function (PDF).

We solve the weighted nonlinear least-squares problem
\[
\min_{\theta}\ \sum_{(u,v)\in\mathcal{A}} w_{uv}\big(\widehat{J}^{\text{emp}}_{uv} - \widehat{J}^{\text{model}}_{uv}(\theta)\big)^2,
\]
subject to
\[
\alpha_0^2 + e_i^\top \Sigma e_j \in (-1,1),\qquad
0 \le \alpha_0^2 + e_i^\top \Sigma e_i < 1.
\]
Weights $w_{uv}$ may be uniform or proportional to effective sample size.

\section{Empirical Analysis}
Having established the modeling framework and estimation procedure, we next take the model to the data and assess its empirical performance. We first describe the data and construct synthetic deal probabilities, then examine the estimated dependence structure through in-sample and out-of-sample fit, and finally analyze the distribution of implied correlations.
\subsection{Data description}
Our dataset consists of 9,255 VC deals with observed outcomes and attribute information. Deals are categorized along three dimensions:
\begin{itemize}[leftmargin=3.6em]
	\item Founder type: first-time vs.\ repeat;
	\item Geography: California, New York, Other U.S., International;
	\item Market category: SaaS, AI, Fintech, Consumer, DevTools, Health.
\end{itemize}

\begin{table}[H]
	\centering
	\caption{Number of Deals by Founder Type, Geography, and Market Category}
	\label{tab:data_1}
	\begin{adjustbox}{max width=\textwidth}
		\begin{tabular}{lcccccccccccc}
			\toprule
			& F$_{\text{first}}$ & F$_{\text{repeat}}$ 
			& G$_{\text{CA}}$ & G$_{\text{NY}}$ & G$_{\text{OtherUS}}$ & G$_{\text{Intl}}$
			& M$_{\text{SaaS}}$ & M$_{\text{AI}}$ & M$_{\text{Fintech}}$ 
			& M$_{\text{Consumer}}$ & M$_{\text{DevTools}}$ & M$_{\text{Health}}$ \\
			\midrule
			Number of Deals
			& 8833 & 422 & 3003 & 1306 & 4845 & 101
			& 4162 & 1986 & 883 & 4294 & 120 & 2108 \\
			\bottomrule
		\end{tabular}
	\end{adjustbox}
\end{table}

\begin{table}[H]
	\centering
	\caption{Number of Distinct Deal Pairs by Attribute Combination ($i \neq j$)}
	\label{tab:data_2}
	\begin{adjustbox}{max width=\textwidth}
		\begin{tabular}{lcccccccccccc}
			\toprule
			& F$_{\text{first}}$ & F$_{\text{repeat}}$ 
			& G$_{\text{CA}}$ & G$_{\text{NY}}$ & G$_{\text{OtherUS}}$ & G$_{\text{Intl}}$
			& M$_{\text{SaaS}}$ & M$_{\text{AI}}$ & M$_{\text{Fintech}}$ 
			& M$_{\text{Consumer}}$ & M$_{\text{DevTools}}$ & M$_{\text{Health}}$ \\
			\midrule
			
			F$_{\text{first}}$
			& 78,013,056 & 3,727,526 
			& 26,522,684 & 11,534,649 & 42,791,213 & 892,036
			& 36,759,003 & 17,540,457 & 7,798,712
			& 37,924,781 & 1,059,852 & 18,617,927 \\
			
			F$_{\text{repeat}}$
			& 3,727,526 & 177,662 
			& 1,267,078 & 551,075 & 2,044,417 & 42,618
			& 1,756,145 & 837,987 & 372,570
			& 1,811,895 & 50,628 & 889,505 \\
			
			G$_{\text{CA}}$
			& 26,522,684 & 1,267,078 
			& 9,015,006 & 3,921,918 & 14,549,535 & 303,303
			& 12,497,028 & 5,963,262 & 2,651,387
			& 12,893,380 & 360,287 & 6,329,762 \\
			
			G$_{\text{NY}}$
			& 11,534,649 & 551,075 
			& 3,921,918 & 1,704,330 & 6,327,570 & 131,906
			& 5,435,051 & 2,593,413 & 1,153,035
			& 5,607,295 & 156,713 & 2,752,830 \\
			
			G$_{\text{OtherUS}}$
			& 42,791,213 & 2,044,417 
			& 14,549,535 & 6,327,570 & 23,469,180 & 489,345
			& 20,162,754 & 9,621,200 & 4,277,690
			& 20,802,356 & 581,360 & 10,211,947 \\
			
			G$_{\text{Intl}}$
			& 892,036 & 42,618 
			& 303,303 & 131,906 & 489,345 & 10,100
			& 420,315 & 200,569 & 89,170
			& 433,645 & 12,120 & 212,893 \\
			
			M$_{\text{SaaS}}$
			& 36,759,003 & 1,756,145 
			& 12,497,028 & 5,435,051 & 20,162,754 & 420,315
			& 17,318,082 & 8,264,816 & 3,674,772
			& 17,870,126 & 499,344 & 8,773,160 \\
			
			M$_{\text{AI}}$
			& 17,540,457 & 837,987 
			& 5,963,262 & 2,593,413 & 9,621,200 & 200,569
			& 8,264,816 & 3,942,210 & 1,753,403
			& 8,527,006 & 238,299 & 4,186,302 \\
			
			M$_{\text{Fintech}}$
			& 7,798,712 & 372,570 
			& 2,651,387 & 1,153,035 & 4,277,690 & 89,170
			& 3,674,772 & 1,753,403 & 778,806
			& 3,791,367 & 105,956 & 1,861,341 \\
			
			M$_{\text{Consumer}}$
			& 37,924,781 & 1,811,895 
			& 12,893,380 & 5,607,295 & 20,802,356 & 433,645
			& 17,870,126 & 8,527,006 & 3,791,367
			& 18,434,142 & 515,229 & 9,051,432 \\
			
			M$_{\text{DevTools}}$
			& 1,059,852 & 50,628 
			& 360,287 & 156,713 & 581,360 & 12,120
			& 499,344 & 238,299 & 105,956
			& 515,229 & 14,280 & 252,959 \\
			
			M$_{\text{Health}}$
			& 18,617,927 & 889,505 
			& 6,329,762 & 2,752,830 & 10,211,947 & 212,893
			& 8,773,160 & 4,186,302 & 1,861,341
			& 9,051,432 & 252,959 & 4,441,556 \\
			
			\bottomrule
		\end{tabular}
	\end{adjustbox}
\end{table}

\subsection{Synthetic deal probability construction}
Each deal is assigned a synthetic deal success probability between 5\% and 20\%, modulated by segment. The construction rule is as follows:

\begin{itemize}[leftmargin=3.6em]
	\item Repeat founders: $p_i \sim U[0.12,0.20]$;
	\item First-time founders: $p_i \sim U[0.05,0.12]$;
	\item Additive nudges (capped at 20\%):\\
	+1\% for CA or NY;
	+1\% for hot sectors (AI, Fintech, SaaS).
\end{itemize}

\paragraph{Selection bias and dataset success rates.} A critical caveat is that the observed success rate in our dataset (approximately 9\% across all deals) substantially exceeds the population-level base rate for early-stage venture investment, which empirical evidence suggests is closer to 1.9\% to 2\% across the broader deal population. This gap is not a modeling error but a well-documented consequence of selection bias. Datasets drawn from VC deals with observable outcomes systematically oversample higher-quality or more visible deals, mechanically inflating apparent success rates. \citet{cochrane2005venture} documents exactly this pattern and develops selection-correction methods to address it.

The synthetic probability construction is designed to reflect the characteristics of the observed, selected sample rather than the unobserved full population. This is a deliberate choice that preserves internal consistency: using the same probability specification for both estimation and simulation ensures that simulated joint success behavior aligns with the observed data from which $\Sigma$ is estimated.

Importantly, the main finding concerns the relative effect of correlation on portfolio outcome distributions: how dependence shifts variance, skewness, and kurtosis while leaving the mean largely unchanged. This comparison between correlated and independent portfolios is structurally robust to the level of marginal success probabilities. The Gaussian copula framework separates the marginal rates (captured by thresholds $t_i = \Phi^{-1}(p_i)$) from the dependence structure $\Sigma$. Scaling marginal probabilities toward population-level rates would compress absolute success counts but preserve the qualitative finding that correlation amplifies tail probabilities relative to independence. The portfolio construction implications, that concentrated and correlated portfolios exhibit heavier right tails at the cost of higher variance, hold across reasonable probability specifications.

We report empirical success frequencies and mean synthetic probabilities by attribute bucket:

\begin{table}[htbp]
\centering
\caption{Empirical and Synthetic Deal Success Probabilities by Attribute}
\label{tab:emp_prob_1}
\begin{adjustbox}{max width=\textwidth}
\begin{tabular}{lcccccccccccc}
\toprule
& F$_{\text{first}}$ & F$_{\text{repeat}}$ 
& G$_{\text{CA}}$ & G$_{\text{NY}}$ & G$_{\text{OtherUS}}$ & G$_{\text{Intl}}$
& M$_{\text{SaaS}}$ & M$_{\text{AI}}$ & M$_{\text{Fintech}}$ 
& M$_{\text{Consumer}}$ & M$_{\text{DevTools}}$ & M$_{\text{Health}}$ \\
\midrule
Number of Deals
& 8,833 & 422 & 3,003 & 1,306 & 4,845 & 101
& 4,162 & 1,986 & 883 & 4,294 & 120 & 2,108 \\
Empirical Success Prob.
& 8.39\% & 32.46\% & 14.45\% & 10.11\% & 6.21\% & 10.89\%
& 7.90\% & 12.08\% & 17.78\% & 5.87\% & 16.67\% & 11.76\% \\
Synthetic Prob.
& 9.68\% & 17.30\% & 10.75\% & 10.46\% & 9.48\% & 9.42\%
& 10.62\% & 10.92\% & 10.84\% & 9.89\% & 10.96\% & 9.33\% \\
\bottomrule
\end{tabular}
\end{adjustbox}
\end{table}

Specifically, we compare these statistics separately for first-time and repeat founders, both with and without conditioning on geography and market category:

\begin{table}[htbp]
\centering
\caption{Empirical and Synthetic Probabilities Under Conditional Buckets}
\label{tab:emp_prob_2}
\begin{adjustbox}{max width=\textwidth}
\begin{tabular}{ccccccc}
\toprule
Constraint
& \multicolumn{2}{c}{\# Deals}
& \multicolumn{2}{c}{Empirical Prob.}
& \multicolumn{2}{c}{Synthetic Prob.} \\
\cmidrule(lr){2-3}\cmidrule(lr){4-5}\cmidrule(lr){6-7}
& First & Repeat & First & Repeat & First & Repeat \\
\midrule
None
& 8,833 & 422 & 8.39\% & 32.46\% & 9.68\% & 17.30\% \\
CA / NY
& 4,064 & 245 & 11.52\% & 40.00\% & 10.25\% & 17.63\% \\
Other US / Intl
& 4,769 & 177 & 5.72\% & 22.03\% & 9.21\% & 16.84\% \\
Hot Sectors
& 5,397 & 299 & 8.78\% & 33.11\% & 10.16\% & 17.61\% \\
Non-Hot Sectors
& 3,436 & 123 & 7.77\% & 30.89\% & 8.93\% & 16.55\% \\
\bottomrule
\end{tabular}
\end{adjustbox}
\end{table}

We use the constructed synthetic probabilities to obtain $(t_{i_k},t_{j_k})$ in Equation \eqref{eq:model_joint_prob_equ} in section 3.6 for the estimation of $\theta=(\alpha_0,\Sigma)$. This implies that the estimated dependence depends on the probability specification. The synthetic construction is designed to preserve salient empirical properties while avoiding degeneracy under fine-grained conditioning, where empirical probabilities can be zero due to sparsity. Using a consistent probability specification between estimation and simulation ensures that simulated joint success behavior aligns with observed data.

\subsection{Estimated dependence structure}
\subsubsection{Training fit}

We estimate the model using 5,000 sampled pairs from the dataset for each attribute combination $(u,v)$. Table~\ref{tab:train_empirical} reports the empirical joint probabilities, while Table~\ref{tab:train_model} reports the model-implied joint probabilities based on the estimated parameters.

\begin{table}[H]
	\centering
	\caption{Empirical Joint Success Probabilities (5,000 Pairs per Attribute Pair)}
	\label{tab:train_empirical}
	\begin{adjustbox}{max width=\textwidth}
		\begin{tabular}{lcccccccccccc}
			\toprule
			& F$_{\text{first}}$ & F$_{\text{repeat}}$ 
			& G$_{\text{CA}}$ & G$_{\text{NY}}$ & G$_{\text{OtherUS}}$ & G$_{\text{Intl}}$
			& M$_{\text{SaaS}}$ & M$_{\text{AI}}$ & M$_{\text{Fintech}}$ 
			& M$_{\text{Consumer}}$ & M$_{\text{DevTools}}$ & M$_{\text{Health}}$ \\
			\midrule
			F$_{\text{first}}$ & 0.60\% & 3.02\% & 1.34\% & 0.74\% & 0.72\% & 0.86\% & 0.66\% & 1.24\% & 1.24\% & 0.32\% & 1.44\% & 0.88\% \\
			F$_{\text{repeat}}$ & 3.02\% & 9.90\% & 4.64\% & 3.14\% & 2.16\% & 3.56\% & 2.96\% & 3.98\% & 5.96\% & 2.06\% & 5.48\% & 3.78\% \\
			G$_{\text{CA}}$ & 1.34\% & 4.64\% & 2.10\% & 1.54\% & 1.04\% & 1.48\% & 0.88\% & 1.66\% & 2.46\% & 0.64\% & 2.40\% & 1.86\% \\
			G$_{\text{NY}}$ & 0.74\% & 3.14\% & 1.54\% & 1.04\% & 0.64\% & 0.78\% & 0.76\% & 1.34\% & 2.06\% & 0.66\% & 1.50\% & 1.28\% \\
			G$_{\text{OtherUS}}$ & 0.72\% & 2.16\% & 1.04\% & 0.64\% & 0.46\% & 0.64\% & 0.46\% & 0.72\% & 0.96\% & 0.34\% & 1.24\% & 0.46\% \\
			G$_{\text{Intl}}$ & 0.86\% & 3.56\% & 1.48\% & 0.78\% & 0.64\% & 1.00\% & 0.98\% & 1.22\% & 1.88\% & 0.72\% & 1.76\% & 1.12\% \\
			M$_{\text{SaaS}}$ & 0.66\% & 2.96\% & 0.88\% & 0.76\% & 0.46\% & 0.98\% & 0.60\% & 1.10\% & 1.32\% & 0.52\% & 1.68\% & 0.94\% \\
			M$_{\text{AI}}$ & 1.24\% & 3.98\% & 1.66\% & 1.34\% & 0.72\% & 1.22\% & 1.10\% & 1.36\% & 2.26\% & 0.52\% & 1.74\% & 1.38\% \\
			M$_{\text{Fintech}}$ & 1.24\% & 5.96\% & 2.46\% & 2.06\% & 0.96\% & 1.88\% & 1.32\% & 2.26\% & 3.16\% & 1.04\% & 3.14\% & 2.28\% \\
			M$_{\text{Consumer}}$ & 0.32\% & 2.06\% & 0.64\% & 0.66\% & 0.34\% & 0.72\% & 0.52\% & 0.52\% & 1.04\% & 0.44\% & 1.08\% & 0.76\% \\
			M$_{\text{DevTools}}$ & 1.44\% & 5.48\% & 2.40\% & 1.50\% & 1.24\% & 1.76\% & 1.68\% & 1.74\% & 3.14\% & 1.08\% & 2.74\% & 2.12\% \\
			M$_{\text{Health}}$ & 0.88\% & 3.78\% & 1.86\% & 1.28\% & 0.46\% & 1.12\% & 0.94\% & 1.38\% & 2.28\% & 0.76\% & 2.12\% & 1.64\% \\
			\bottomrule
		\end{tabular}
	\end{adjustbox}
\end{table}

\begin{table}[H]
	\centering
	\caption{Model-Implied Joint Success Probabilities (5,000 Pairs per Attribute Pair)}
	\label{tab:train_model}
	\begin{adjustbox}{max width=\textwidth}
		\begin{tabular}{lcccccccccccc}
			\toprule
			& F$_{\text{first}}$ & F$_{\text{repeat}}$ 
			& G$_{\text{CA}}$ & G$_{\text{NY}}$ & G$_{\text{OtherUS}}$ & G$_{\text{Intl}}$
			& M$_{\text{SaaS}}$ & M$_{\text{AI}}$ & M$_{\text{Fintech}}$ 
			& M$_{\text{Consumer}}$ & M$_{\text{DevTools}}$ & M$_{\text{Health}}$ \\
			\midrule
			F$_{\text{first}}$ & 0.94\% & 1.62\% & 1.02\% & 1.01\% & 0.91\% & 0.91\% & 1.01\% & 1.05\% & 1.03\% & 0.95\% & 1.05\% & 0.90\% \\
			F$_{\text{repeat}}$ & 1.62\% & 4.38\% & 2.45\% & 1.88\% & 1.82\% & 1.65\% & 2.23\% & 2.10\% & 2.31\% & 1.84\% & 2.13\% & 1.79\% \\
			G$_{\text{CA}}$ & 1.02\% & 2.45\% & 1.41\% & 1.16\% & 1.10\% & 1.02\% & 1.31\% & 1.27\% & 1.35\% & 1.12\% & 1.28\% & 1.08\% \\
			G$_{\text{NY}}$ & 1.01\% & 1.88\% & 1.16\% & 1.10\% & 1.00\% & 0.99\% & 1.13\% & 1.15\% & 1.16\% & 1.04\% & 1.16\% & 0.99\% \\
			G$_{\text{OtherUS}}$ & 0.91\% & 1.82\% & 1.10\% & 1.00\% & 0.92\% & 0.90\% & 1.06\% & 1.06\% & 1.08\% & 0.95\% & 1.07\% & 0.91\% \\
			G$_{\text{Intl}}$ & 0.91\% & 1.65\% & 1.02\% & 0.99\% & 0.90\% & 0.89\% & 1.01\% & 1.03\% & 1.03\% & 0.93\% & 1.04\% & 0.88\% \\
			M$_{\text{SaaS}}$ & 1.01\% & 2.23\% & 1.31\% & 1.13\% & 1.06\% & 1.01\% & 1.25\% & 1.22\% & 1.28\% & 1.09\% & 1.23\% & 1.04\% \\
			M$_{\text{AI}}$ & 1.05\% & 2.10\% & 1.27\% & 1.15\% & 1.06\% & 1.03\% & 1.22\% & 1.23\% & 1.25\% & 1.10\% & 1.23\% & 1.05\% \\
			M$_{\text{Fintech}}$ & 1.03\% & 2.31\% & 1.35\% & 1.16\% & 1.08\% & 1.03\% & 1.28\% & 1.25\% & 1.32\% & 1.11\% & 1.27\% & 1.07\% \\
			M$_{\text{Consumer}}$ & 0.95\% & 1.84\% & 1.12\% & 1.04\% & 0.95\% & 0.93\% & 1.09\% & 1.10\% & 1.11\% & 0.99\% & 1.11\% & 0.94\% \\
			M$_{\text{DevTools}}$ & 1.05\% & 2.13\% & 1.28\% & 1.16\% & 1.07\% & 1.04\% & 1.23\% & 1.23\% & 1.27\% & 1.11\% & 1.24\% & 1.05\% \\
			M$_{\text{Health}}$ & 0.90\% & 1.79\% & 1.08\% & 0.99\% & 0.91\% & 0.88\% & 1.04\% & 1.05\% & 1.07\% & 0.94\% & 1.05\% & 0.89\% \\
			\bottomrule
		\end{tabular}
	\end{adjustbox}
\end{table}

The goodness-of-fit between empirical and model-implied joint probabilities is summarized by the mean squared error (MSE) and root mean squared error (RMSE):

\[
\text{MSE} = 1.13 \times 10^{-4}, 
\qquad
\text{RMSE} = 0.010619 \; (1.0619\%).
\]

The relatively small RMSE suggests that the model captures broad dependence patterns across attribute pairs.

\subsubsection{Full-sample robustness check}   

To assess estimation stability, we evaluate model performance using all available sample pairs for each $(u, v)$ combination. We emphasize that this constitutes a robustness check on parameter stability rather than a true out-of-sample validation: the full-sample pairs are drawn from the same underlying dataset used in estimation, not from a temporally or geographically distinct holdout. A genuine out-of-sample evaluation would require a vintage-based temporal split or a separate dataset, which we leave for future work. Table~\ref{tab:robust_empirical} reports empirical joint probabilities computed using the full sample, and Table~\ref{tab:robust_model} reports the corresponding model-implied probabilities based on the previously estimated parameters.

\begin{table}[H]
	\centering
	\caption{Empirical Joint Success Probabilities (All Sample Pairs)}
	\label{tab:robust_empirical}
	\begin{adjustbox}{max width=\textwidth}
		\begin{tabular}{lcccccccccccc}
			\toprule
			& F$_{\text{first}}$ & F$_{\text{repeat}}$ 
			& G$_{\text{CA}}$ & G$_{\text{NY}}$ & G$_{\text{OtherUS}}$ & G$_{\text{Intl}}$
			& M$_{\text{SaaS}}$ & M$_{\text{AI}}$ & M$_{\text{Fintech}}$ 
			& M$_{\text{Consumer}}$ & M$_{\text{DevTools}}$ & M$_{\text{Health}}$ \\
			\midrule
			F$_{\text{first}}$ & 0.70\% & 2.72\% & 1.21\% & 0.85\% & 0.52\% & 0.91\% & 0.66\% & 1.01\% & 1.49\% & 0.49\% & 1.40\% & 0.99\% \\
			F$_{\text{repeat}}$ & 2.72\% & 10.49\% & 4.69\% & 3.28\% & 2.02\% & 3.54\% & 2.56\% & 3.92\% & 5.77\% & 1.90\% & 5.41\% & 3.82\% \\
			G$_{\text{CA}}$ & 1.21\% & 4.69\% & 2.08\% & 1.46\% & 0.90\% & 1.57\% & 1.14\% & 1.74\% & 2.57\% & 0.85\% & 2.40\% & 1.70\% \\
			G$_{\text{NY}}$ & 0.85\% & 3.28\% & 1.46\% & 1.01\% & 0.63\% & 1.10\% & 0.80\% & 1.22\% & 1.79\% & 0.59\% & 1.68\% & 1.19\% \\
			G$_{\text{OtherUS}}$ & 0.52\% & 2.02\% & 0.90\% & 0.63\% & 0.38\% & 0.68\% & 0.49\% & 0.75\% & 1.10\% & 0.36\% & 1.04\% & 0.73\% \\
			G$_{\text{Intl}}$ & 0.91\% & 3.54\% & 1.57\% & 1.10\% & 0.68\% & 1.09\% & 0.86\% & 1.32\% & 1.94\% & 0.64\% & 1.82\% & 1.28\% \\
			M$_{\text{SaaS}}$ & 0.66\% & 2.56\% & 1.14\% & 0.80\% & 0.49\% & 0.86\% & 0.62\% & 0.95\% & 1.40\% & 0.46\% & 1.31\% & 0.93\% \\
			M$_{\text{AI}}$ & 1.01\% & 3.92\% & 1.74\% & 1.22\% & 0.75\% & 1.32\% & 0.95\% & 1.46\% & 2.15\% & 0.71\% & 2.01\% & 1.42\% \\
			M$_{\text{Fintech}}$ & 1.49\% & 5.77\% & 2.57\% & 1.79\% & 1.10\% & 1.94\% & 1.40\% & 2.15\% & 3.14\% & 1.04\% & 2.96\% & 2.09\% \\
			M$_{\text{Consumer}}$ & 0.49\% & 1.90\% & 0.85\% & 0.59\% & 0.36\% & 0.64\% & 0.46\% & 0.71\% & 1.04\% & 0.34\% & 0.98\% & 0.69\% \\
			M$_{\text{DevTools}}$ & 1.40\% & 5.41\% & 2.40\% & 1.68\% & 1.04\% & 1.82\% & 1.31\% & 2.01\% & 2.96\% & 0.98\% & 2.66\% & 1.96\% \\
			M$_{\text{Health}}$ & 0.99\% & 3.82\% & 1.70\% & 1.19\% & 0.73\% & 1.28\% & 0.93\% & 1.42\% & 2.09\% & 0.69\% & 1.96\% & 1.38\% \\
			\bottomrule
		\end{tabular}
	\end{adjustbox}
\end{table}

\begin{table}[H]
	\centering
	\caption{Model-Implied Joint Success Probabilities (All Sample Pairs)}
	\label{tab:robust_model}
	\begin{adjustbox}{max width=\textwidth}
		\begin{tabular}{lcccccccccccc}
			\toprule
			& F$_{\text{first}}$ & F$_{\text{repeat}}$ 
			& G$_{\text{CA}}$ & G$_{\text{NY}}$ & G$_{\text{OtherUS}}$ & G$_{\text{Intl}}$
			& M$_{\text{SaaS}}$ & M$_{\text{AI}}$ & M$_{\text{Fintech}}$ 
			& M$_{\text{Consumer}}$ & M$_{\text{DevTools}}$ & M$_{\text{Health}}$ \\
			\midrule
			F$_{\text{first}}$ & 1.18\% & 2.67\% & 1.46\% & 1.20\% & 1.13\% & 1.07\% & 1.39\% & 1.42\% & 1.46\% & 1.23\% & 1.56\% & 1.12\% \\
			F$_{\text{repeat}}$ & 2.67\% & 7.09\% & 3.57\% & 2.57\% & 2.53\% & 2.27\% & 3.33\% & 3.39\% & 3.55\% & 2.81\% & 3.93\% & 2.52\% \\
			G$_{\text{CA}}$ & 1.46\% & 3.57\% & 1.88\% & 1.44\% & 1.39\% & 1.28\% & 1.77\% & 1.80\% & 1.87\% & 1.53\% & 2.04\% & 1.38\% \\
			G$_{\text{NY}}$ & 1.20\% & 2.57\% & 1.44\% & 1.24\% & 1.16\% & 1.11\% & 1.39\% & 1.42\% & 1.45\% & 1.24\% & 1.53\% & 1.14\% \\
			G$_{\text{OtherUS}}$ & 1.13\% & 2.53\% & 1.39\% & 1.16\% & 1.09\% & 1.03\% & 1.33\% & 1.36\% & 1.39\% & 1.18\% & 1.48\% & 1.08\% \\
			G$_{\text{Intl}}$ & 1.07\% & 2.27\% & 1.28\% & 1.11\% & 1.03\% & 0.99\% & 1.24\% & 1.27\% & 1.28\% & 1.11\% & 1.36\% & 1.02\% \\
			M$_{\text{SaaS}}$ & 1.39\% & 3.33\% & 1.77\% & 1.39\% & 1.33\% & 1.24\% & 1.67\% & 1.71\% & 1.76\% & 1.45\% & 1.91\% & 1.32\% \\
			M$_{\text{AI}}$ & 1.42\% & 3.39\% & 1.80\% & 1.42\% & 1.36\% & 1.27\% & 1.71\% & 1.74\% & 1.80\% & 1.48\% & 1.95\% & 1.35\% \\
			M$_{\text{Fintech}}$ & 1.46\% & 3.55\% & 1.87\% & 1.45\% & 1.39\% & 1.28\% & 1.76\% & 1.80\% & 1.86\% & 1.52\% & 2.02\% & 1.38\% \\
			M$_{\text{Consumer}}$ & 1.23\% & 2.81\% & 1.53\% & 1.24\% & 1.18\% & 1.11\% & 1.45\% & 1.48\% & 1.52\% & 1.28\% & 1.64\% & 1.17\% \\
			M$_{\text{DevTools}}$ & 1.56\% & 3.93\% & 2.04\% & 1.53\% & 1.48\% & 1.36\% & 1.91\% & 1.95\% & 2.02\% & 1.64\% & 2.22\% & 1.47\% \\
			M$_{\text{Health}}$ & 1.12\% & 2.52\% & 1.38\% & 1.14\% & 1.08\% & 1.02\% & 1.32\% & 1.35\% & 1.38\% & 1.17\% & 1.47\% & 1.07\% \\
			\bottomrule
		\end{tabular}
	\end{adjustbox}
\end{table}

The out-of-sample fit metrics are:

\[
\text{MSE} = 5.01 \times 10^{-5}, 
\qquad
\text{RMSE} = 0.007081 \; (0.7081\%).
\]

The lower error relative to the 5,000-pair training fit reflects reduced sampling noise at larger pair counts rather than evidence of external predictive validity. The consistency of parameter estimates across sample sizes suggests that $\Sigma$ is not overfitted to the specific 5,000-pair training sample.

\subsubsection{Estimated covariance structure}

The estimated intercept parameter is
$
\hat{\alpha}_0 = 0.0.
$
The estimated covariance matrix of the latent Gaussian attribute vector,
$
\hat{\Sigma} = \mathrm{Var}(F),
$
is reported in Table~\ref{tab:sigma_est}.

\paragraph{Remark on $\hat{\alpha}_0 = 0$.} This boundary result requires careful interpretation. Examining Equation \eqref{eq:cov}, the global factor contribution $\alpha_0^2$ adds a constant shift to all pairwise latent covariances regardless of deal attributes. Given the flexibility of a full 12-dimensional $\Sigma$ matrix, this constant shift can be absorbed into the elements of $\Sigma$ without changing the value of the objective function. The two components are therefore not separately identified within this framework: a range of ($\alpha_0$, $\Sigma$) pairs can produce equivalent fits to the observed joint frequencies. The result $\hat{\alpha}_0 = 0$ should be read as the optimizer finding a minimum at the boundary of the feasible region, not as a substantive empirical claim that macro-level common shocks are absent in venture investing. Indeed, vintage effects, interest rate environments, and broad technology cycles are well-documented sources of cross-deal dependence that operate independently of founder, geography, and market attributes. Identifying these global factors within this framework would require either richer time-series variation or externally specified macro covariates.

\begin{table}[H]
	\centering
	\caption{Estimated Covariance Matrix $\Sigma$ of Latent Gaussian Attributes}
	\label{tab:sigma_est}
	\begin{adjustbox}{max width=\textwidth}
		\begin{tabular}{lcccccccccccc}
			\toprule
			& F$_{\text{first}}$ & F$_{\text{repeat}}$ 
			& G$_{\text{CA}}$ & G$_{\text{NY}}$ & G$_{\text{OtherUS}}$ & G$_{\text{Intl}}$
			& M$_{\text{SaaS}}$ & M$_{\text{AI}}$ & M$_{\text{Fintech}}$ 
			& M$_{\text{Consumer}}$ & M$_{\text{DevTools}}$ & M$_{\text{Health}}$ \\
			\midrule
			F$_{\text{first}}$
			& 0.0012 & -0.0134 & -0.0081 & -0.0009 & -0.0028 & -0.0003 
			& -0.0051 & -0.0030 & -0.0060 & -0.0017 & -0.0032 & -0.0028 \\
			
			F$_{\text{repeat}}$
			& -0.0134 & 0.1950 & 0.1171 & 0.0152 & 0.0398 & 0.0055 
			& 0.0809 & 0.0440 & 0.0877 & 0.0279 & 0.0482 & 0.0404 \\
			
			G$_{\text{CA}}$
			& -0.0081 & 0.1171 & 0.0713 & 0.0090 & 0.0239 & 0.0036 
			& 0.0489 & 0.0259 & 0.0523 & 0.0172 & 0.0291 & 0.0239 \\
			
			G$_{\text{NY}}$
			& -0.0009 & 0.0152 & 0.0090 & 0.0015 & 0.0031 & 0.0005 
			& 0.0066 & 0.0035 & 0.0069 & 0.0023 & 0.0036 & 0.0030 \\
			
			G$_{\text{OtherUS}}$
			& -0.0028 & 0.0398 & 0.0239 & 0.0031 & 0.0082 & 0.0011 
			& 0.0165 & 0.0090 & 0.0179 & 0.0056 & 0.0098 & 0.0082 \\
			
			G$_{\text{Intl}}$
			& -0.0003 & 0.0055 & 0.0036 & 0.0005 & 0.0011 & 0.0003 
			& 0.0026 & 0.0012 & 0.0024 & 0.0010 & 0.0014 & 0.0010 \\
			
			M$_{\text{SaaS}}$
			& -0.0051 & 0.0809 & 0.0489 & 0.0066 & 0.0165 & 0.0026 
			& 0.0347 & 0.0183 & 0.0363 & 0.0121 & 0.0202 & 0.0165 \\
			
			M$_{\text{AI}}$
			& -0.0030 & 0.0440 & 0.0259 & 0.0035 & 0.0090 & 0.0012 
			& 0.0183 & 0.0107 & 0.0201 & 0.0059 & 0.0109 & 0.0094 \\
			
			M$_{\text{Fintech}}$
			& -0.0060 & 0.0877 & 0.0523 & 0.0069 & 0.0179 & 0.0024 
			& 0.0363 & 0.0201 & 0.0397 & 0.0124 & 0.0217 & 0.0184 \\
			
			M$_{\text{Consumer}}$
			& -0.0017 & 0.0279 & 0.0172 & 0.0023 & 0.0056 & 0.0010 
			& 0.0121 & 0.0059 & 0.0124 & 0.0045 & 0.0070 & 0.0055 \\
			
			M$_{\text{DevTools}}$
			& -0.0032 & 0.0482 & 0.0291 & 0.0036 & 0.0098 & 0.0014 
			& 0.0202 & 0.0109 & 0.0217 & 0.0070 & 0.0124 & 0.0101 \\
			
			M$_{\text{Health}}$
			& -0.0028 & 0.0404 & 0.0239 & 0.0030 & 0.0082 & 0.0010 
			& 0.0165 & 0.0094 & 0.0184 & 0.0055 & 0.0101 & 0.0088 \\
			
			\bottomrule
		\end{tabular}
	\end{adjustbox}
\end{table}

%
%

A notable feature of $\hat{\Sigma}$ is the presence of negative values throughout the F$_{\text{first}}$ row and column. Two factors contribute to this pattern and should be distinguished. First, as noted in Section 3.1, the sum-to-one constraint on the founder type one-hot vector (F$_{\text{first}}$ + F$_{\text{repeat}}$ = 1 for every deal) introduces a mechanical negative relationship between the two founder-type indicators. The optimizer partially absorbs this constraint into the off-diagonal elements of $\Sigma$, producing negative cross-terms that are partly algebraic rather than purely economic in origin. Second, there is a genuine economic component: the performance of first-time founder deals exhibits weaker co-movement with repeat-founder deals due to differences in experience, network access, and information environment. Disentangling these two sources would require dropping the redundant reference category, which we leave for future work.

Importantly, these negative values operate at the level of latent attribute factors. At the level of observable binary outcomes, the implied Bernoulli correlations between deal success indicators remain predominantly positive across all attribute pairs, consistent with economic intuition.

\subsection{Correlation distributions}
\begin{figure}[H]
\centering
\includegraphics[width=0.95\textwidth]{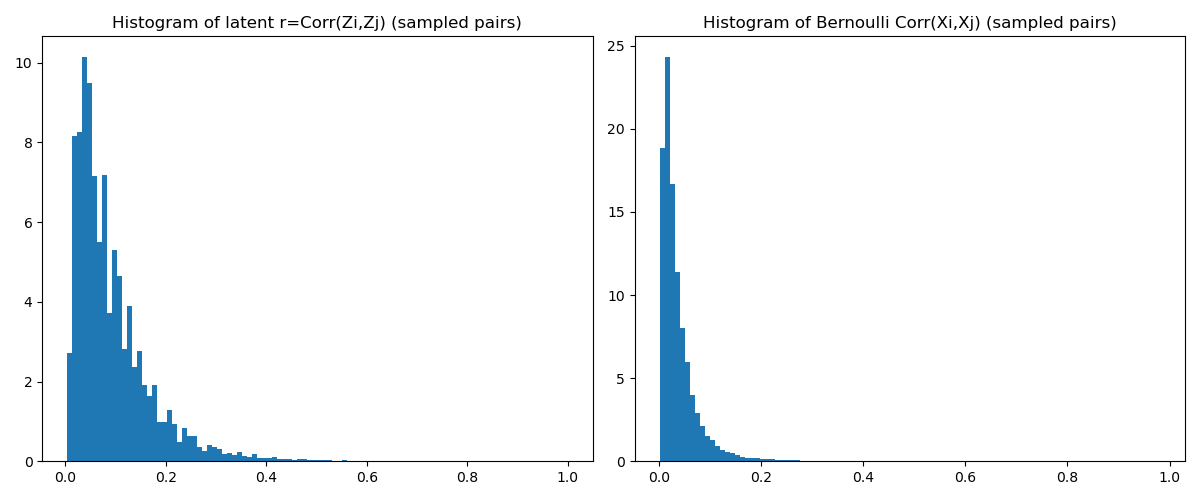}
\caption{Histograms of latent correlations $r_{ij}=\Corr(Z_i,Z_j)$ and induced Bernoulli correlations $\Corr(X_i,X_j)$ across 100,000 sampled deal pairs. }
\label{fig:corr_distributions}
\end{figure}

Figure~\ref{fig:corr_distributions} shows latent vs.\ induced Bernoulli correlations.

Observable correlations are compressed toward zero, yet latent correlations exhibit substantial dispersion. This latent dependence is sufficient to generate pronounced tail amplification at the portfolio level.

\section{Portfolio Simulations}
Building on the estimated dependence structure, we examine the portfolio-level implications through simulation, focusing on the distribution of successful deals. In particular, we analyze how portfolio construction choices interact with correlated outcomes in venture investing by comparing portfolios under different concentration profiles:
\begin{itemize}[leftmargin=3.6em]
  \item \textbf{Portfolio A}: 50\% first-time / 50\% repeat founders, geographically diversified;
  \item \textbf{Portfolio B}: all repeat founders in California;
  \item \textbf{Portfolio C (diversified)}: spread across markets;
  \item \textbf{Portfolio C (concentrated)}: all one-market-labeled deals (e.g., AI-only).
\end{itemize}
We evaluate portfolio sizes $n\in\{20,40,80\}$ via Monte Carlo simulation (50,000 replications for reported summaries). We focus primarily on the distributional analysis of the number of successful deals in the 40-deal portfolio. Full distributional results for the 20- and 80-deal portfolios are reported in Appendix~\ref{app:dist}.

\subsection{Distribution of successful deals}

\begin{figure}[H]
	\centering
	\begin{subfigure}{0.32\textwidth}
		\includegraphics[width=\textwidth]{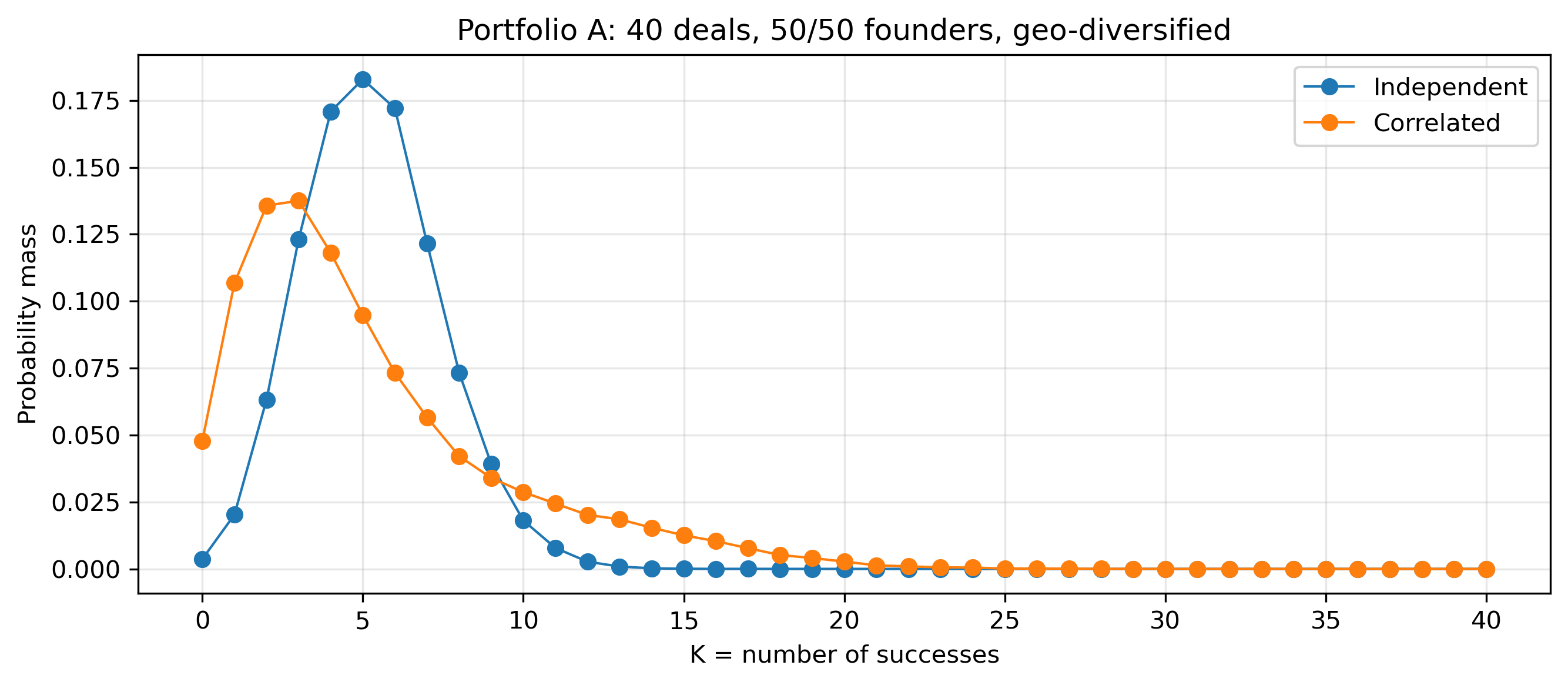}
		\caption{Portfolio A}
	\end{subfigure}
	\hfill
	\begin{subfigure}{0.32\textwidth}
		\includegraphics[width=\textwidth]{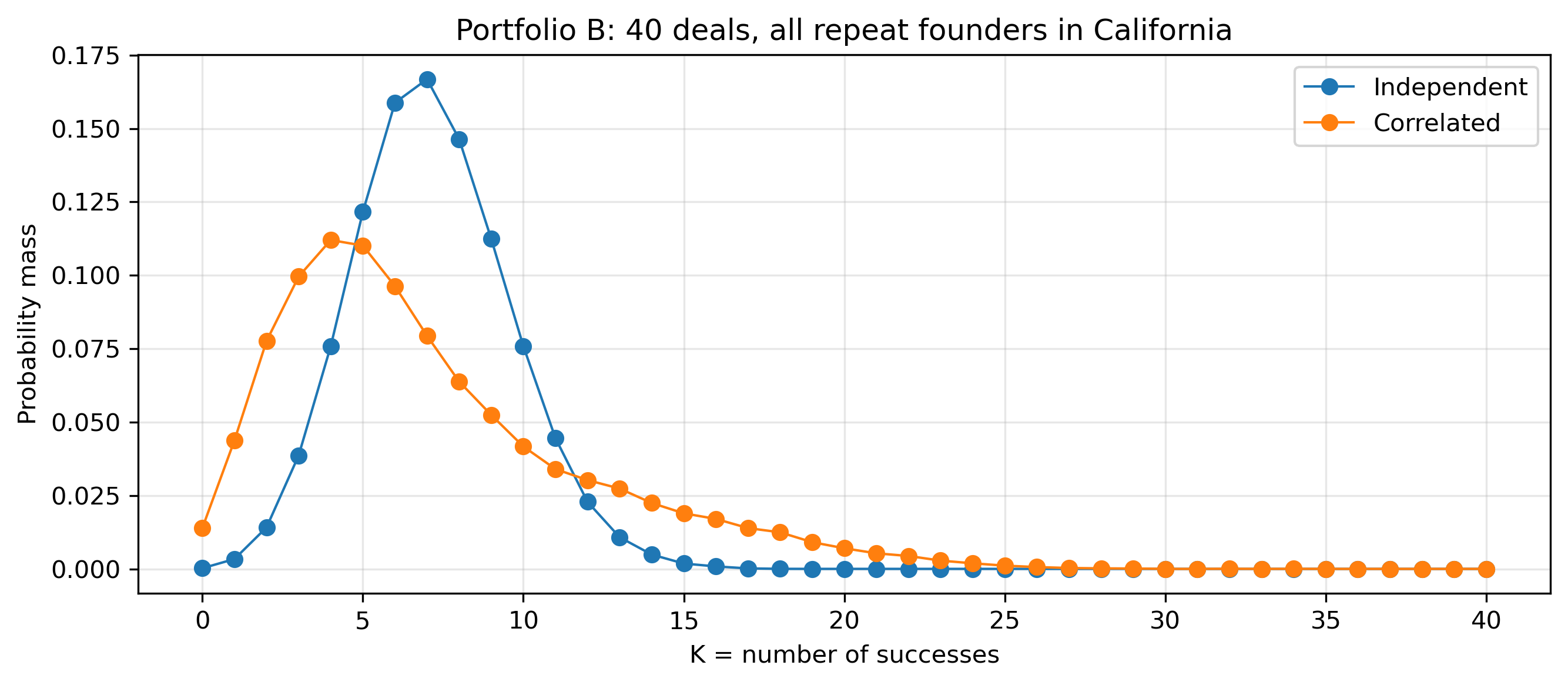}
		\caption{Portfolio B}
	\end{subfigure}
	\hfill
	\begin{subfigure}{0.32\textwidth}
		\includegraphics[width=\textwidth]{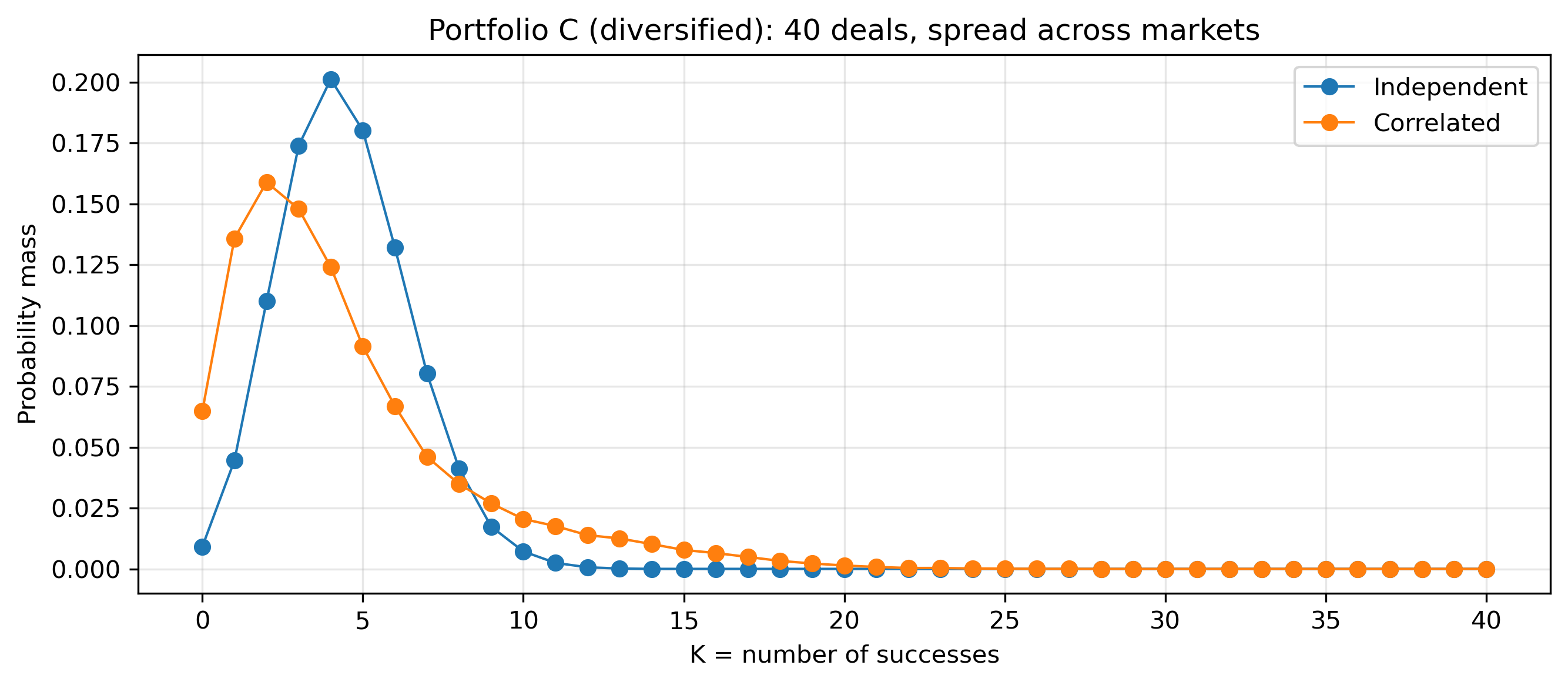}
		\caption{Portfolio C (diversified)}
	\end{subfigure}
	\caption{Distribution of the number of successful deals for Portfolio A (50/50 founders, geographically diversified), Portfolio B (all repeat founders in California), and diversified Portfolio C (spread across markets) with 40 deals.}
	\label{fig:portfolio_ABCdiv_40deals}
\end{figure}

Figure \ref{fig:portfolio_ABCdiv_40deals} illustrates the distribution of the number of successful deals for 40-deal portfolios. Under independence, the distribution of successful deals is comparatively tight across designs. Under correlation, the right tail becomes noticeably heavier, especially for the more concentrated portfolio, producing greater variability. The diversified Portfolio C distribution still shows tail thickening under correlation, but the effect is attenuated relative to Portfolio B, underscoring that market diversification mitigates but does not eliminate correlation-driven tail amplification.

\begin{figure}[H]
	\centering
	\begin{subfigure}{0.32\textwidth}
		\includegraphics[width=\textwidth]{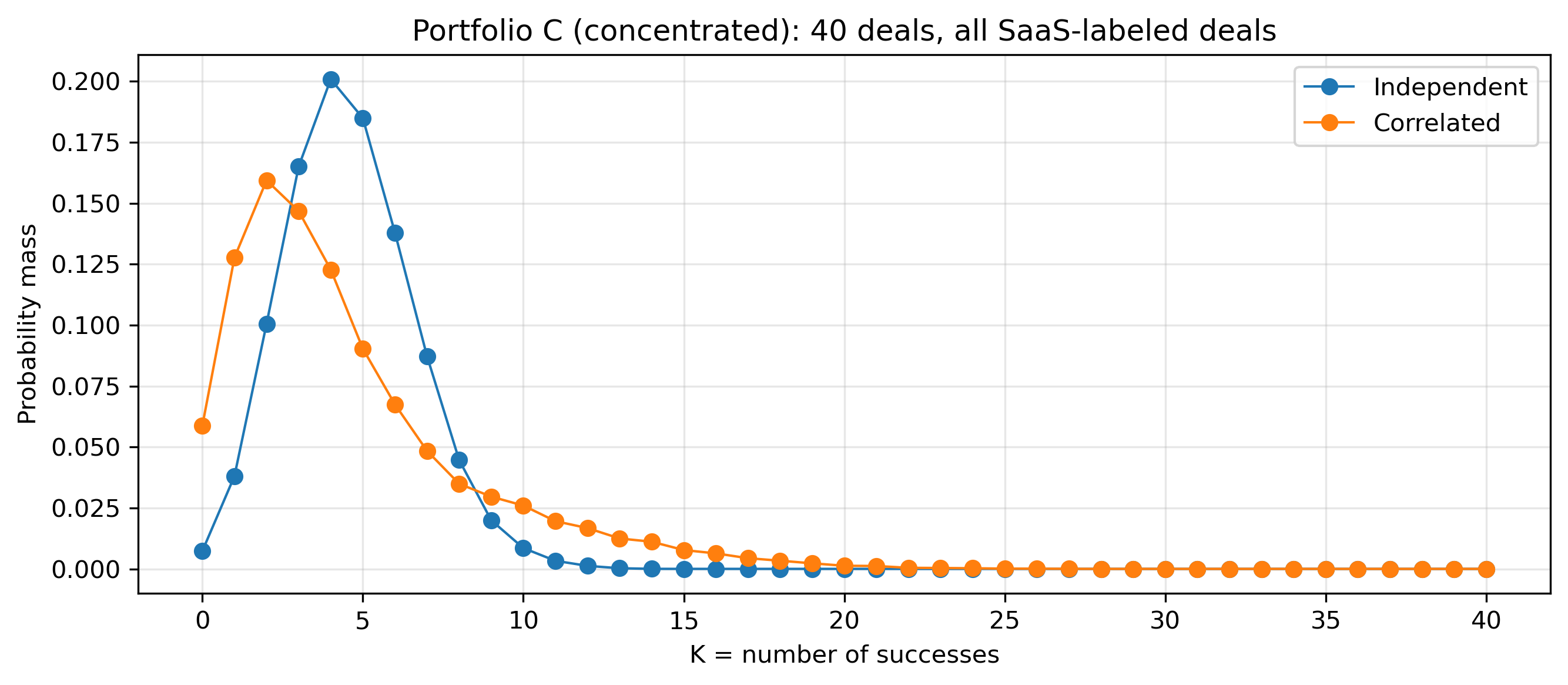}
		\caption{SaaS}
	\end{subfigure}
	\hfill
	\begin{subfigure}{0.32\textwidth}
		\includegraphics[width=\textwidth]{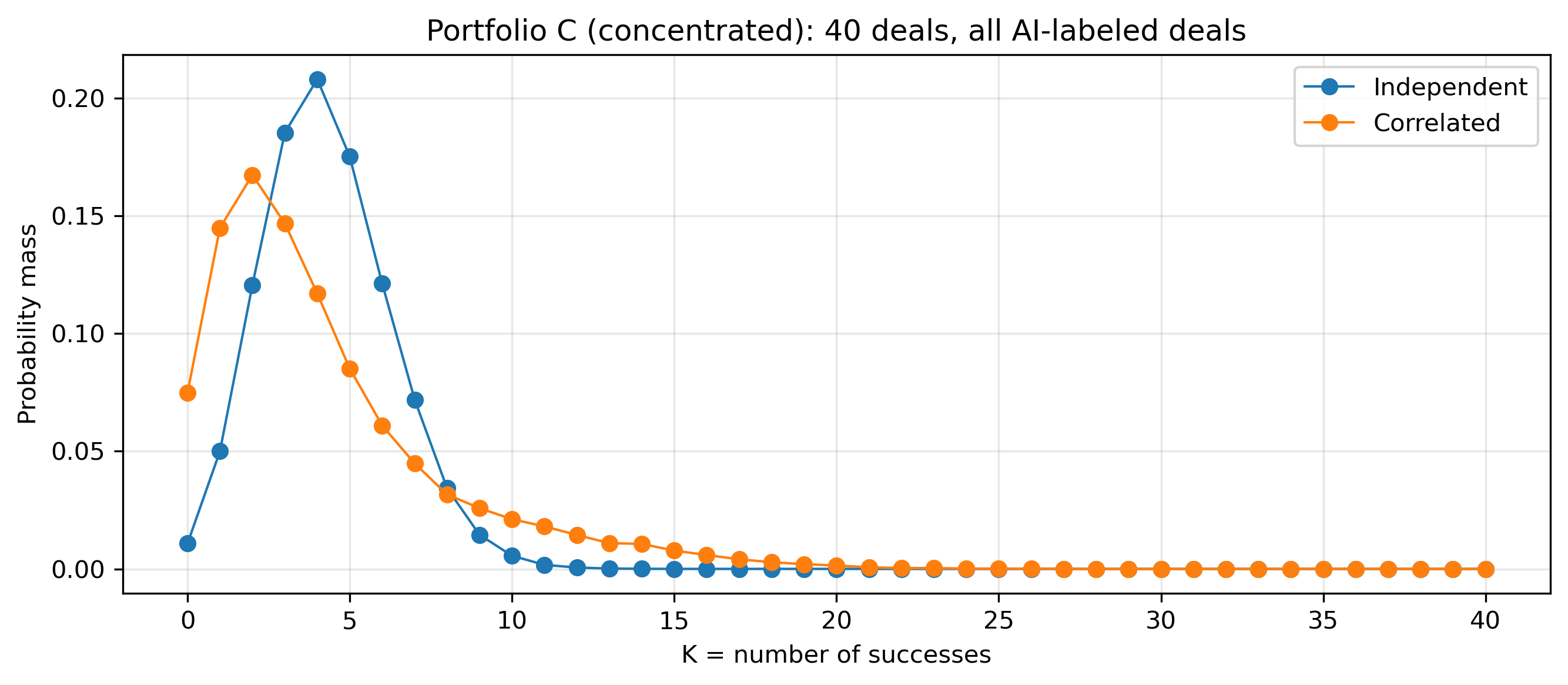}
		\caption{AI}
	\end{subfigure}
	\hfill
	\begin{subfigure}{0.32\textwidth}
		\includegraphics[width=\textwidth]{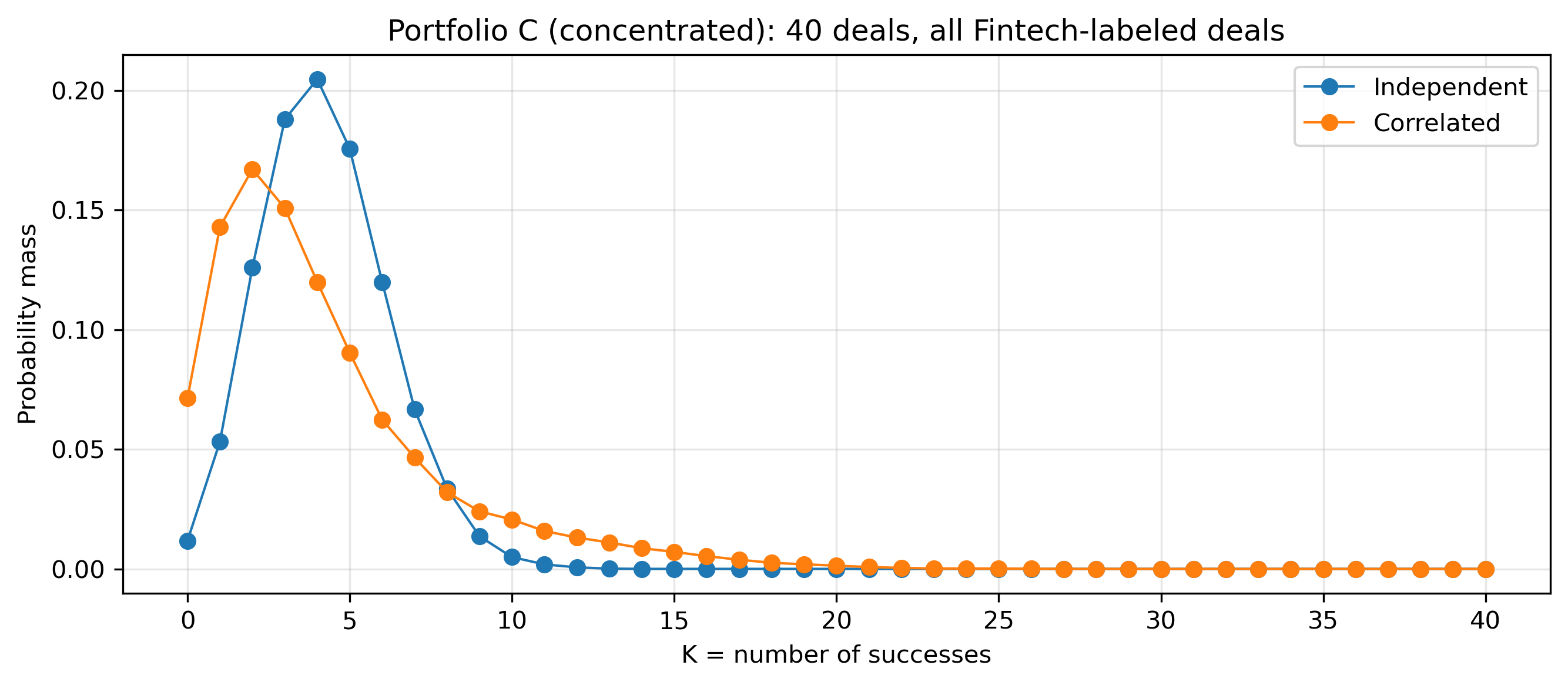}
		\caption{Fintech}
	\end{subfigure}
	
	\vspace{0.5em}
	
	\begin{subfigure}{0.32\textwidth}
		\includegraphics[width=\textwidth]{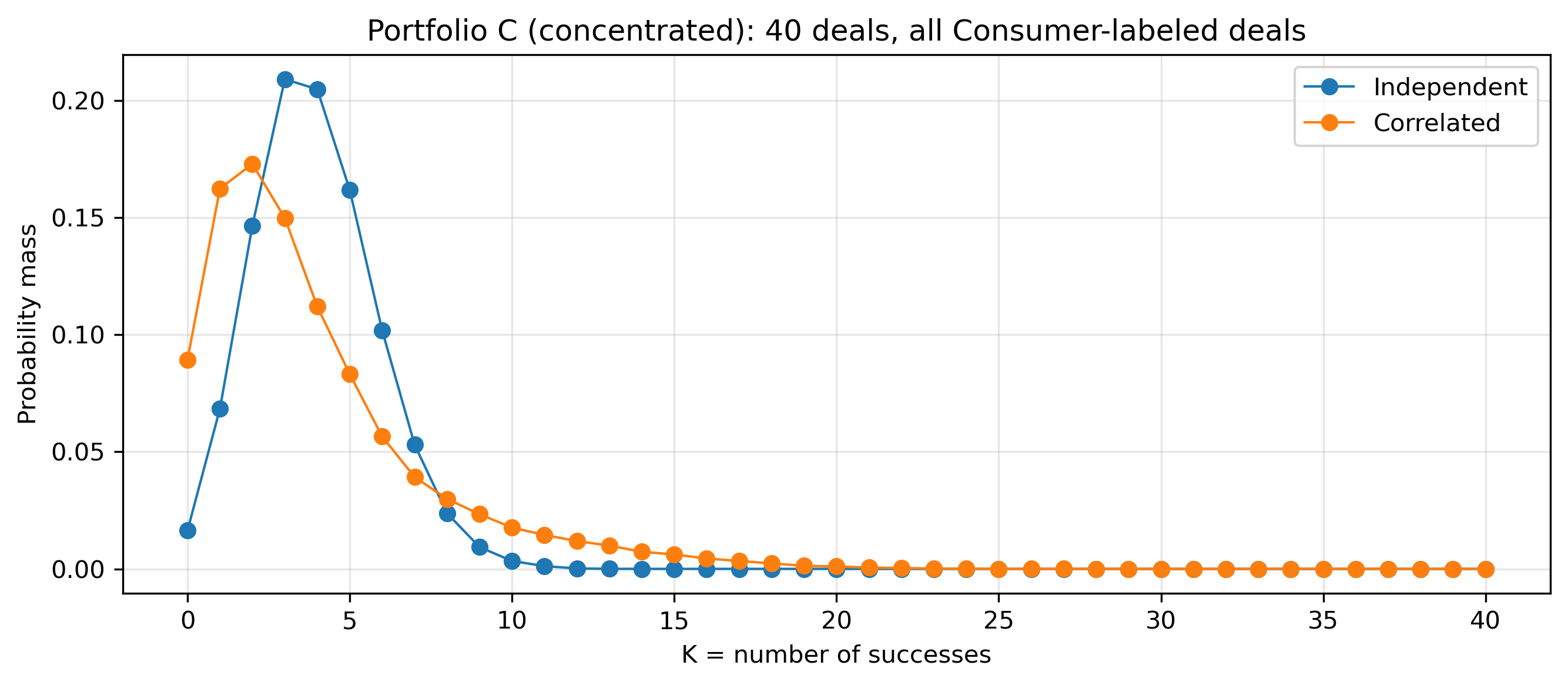}
		\caption{Consumer}
	\end{subfigure}
	\hfill
	\begin{subfigure}{0.32\textwidth}
		\includegraphics[width=\textwidth]{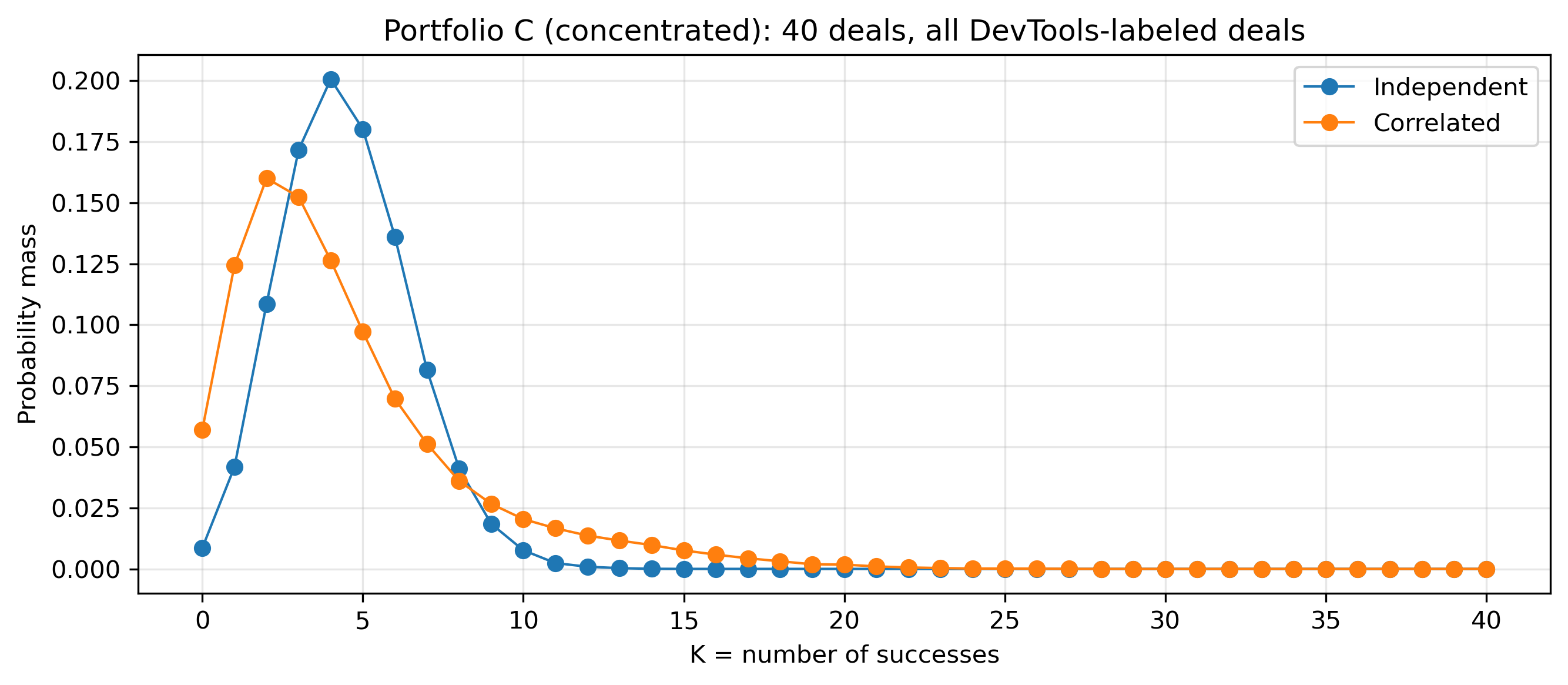}
		\caption{DevTools}
	\end{subfigure}
	\hfill
	\begin{subfigure}{0.32\textwidth}
		\includegraphics[width=\textwidth]{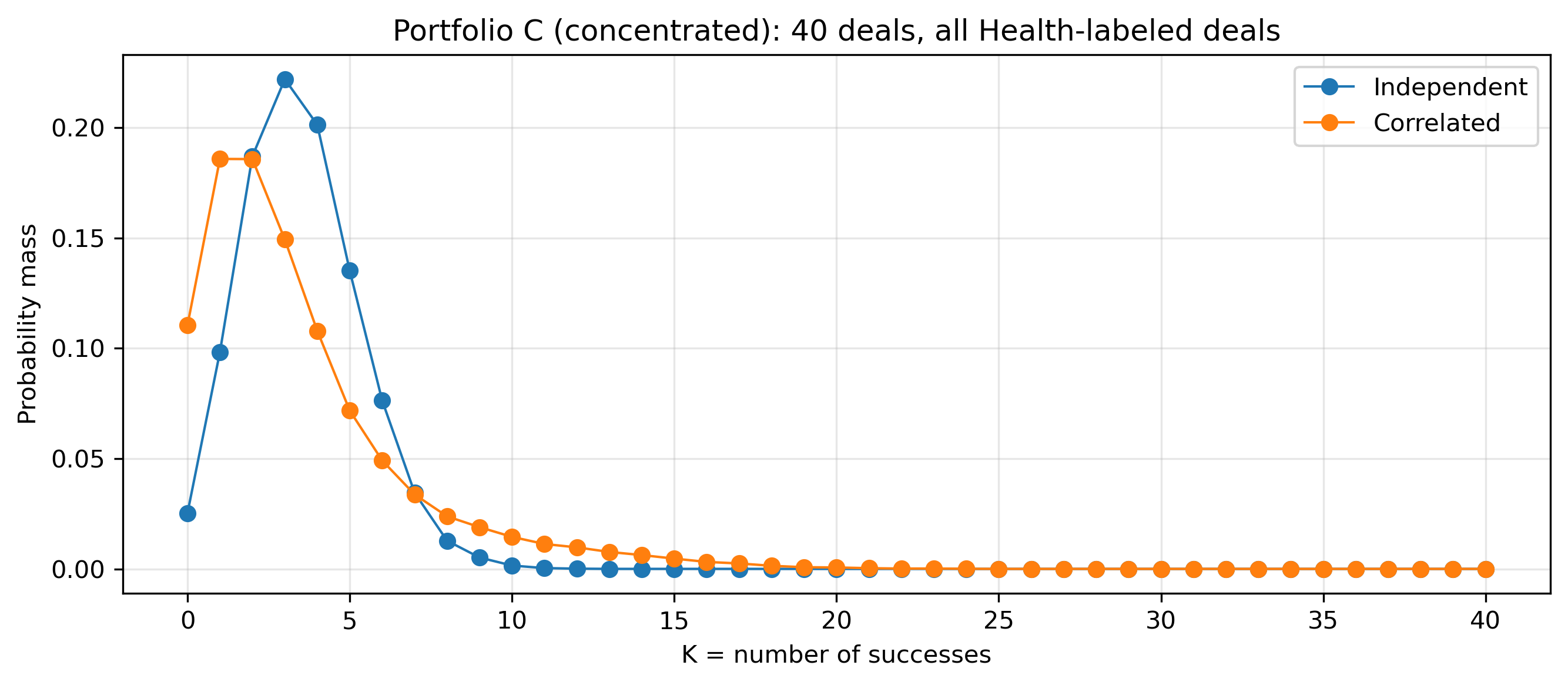}
		\caption{Health}
	\end{subfigure}
	\caption{Concentrated Portfolio C with 40 deals across market categories.}
	\label{fig:portfolio_C_40}
\end{figure}

We then examine concentrated versions of Portfolio C to isolate the role of market-category exposure. Figure \ref{fig:portfolio_C_40} compares 40-deal portfolios concentrated entirely within a single category (SaaS, AI, Fintech, Consumer, DevTools, Health).

%
%
%

Table \ref{tab:K_summary_40} summarizes the distributional moments of the number of successful deals K in the 40-deal case. Across all portfolios, correlation does not materially change expected number of successes because marginal success probabilities are fixed. In contrast, variance, skewness, and kurtosis increase systematically under correlation, indicating substantially heavier right tails. Concentrated portfolios and narrow sector exposures exhibit the strongest tail amplification (e.g., Portfolio B).

Table~\ref{tab:tail_40_full} reports tail probabilities for 40-deal portfolios. Correlation reduces the probability of achieving a small number of successes while sharply amplifying extreme upper-tail events (e.g., $P(K\ge10)$).


\begin{table}[H]
	\centering
	\caption{Summary Statistics of the Number of Successful Deals ($K$), 40-Deal Portfolios}
	\label{tab:K_summary_40}
	\begin{adjustbox}{max width=\textwidth}
		\begin{tabular}{l l c c c c}
			\toprule
			Portfolio & Setting & Mean & Std & Skew & Kurt \\
			\midrule
			Portfolio A & Independent & 5.25 & 2.12 & 0.34 & 3.08 \\
			& Correlated  & 5.26 & 4.24 & 1.31 & 4.58 \\
			Portfolio B & Independent & 7.14 & 2.41 & 0.28 & 3.07 \\
			& Correlated  & 7.15 & 4.85 & 1.12 & 4.02 \\
			Portfolio C (Diversified) & Independent & 4.44 & 1.98 & 0.38 & 3.06 \\
			& Correlated  & 4.44 & 3.77 & 1.54 & 5.64 \\
			Portfolio C -- SaaS & Independent & 4.58 & 2.00 & 0.40 & 3.16 \\
			& Correlated  & 4.59 & 3.82 & 1.47 & 5.40 \\
			Portfolio C -- AI & Independent & 4.28 & 1.94 & 0.41 & 3.15 \\
			& Correlated  & 4.29 & 3.75 & 1.55 & 5.66 \\
			Portfolio C -- Fintech & Independent & 4.22 & 1.94 & 0.41 & 3.15 \\
			& Correlated  & 4.24 & 3.64 & 1.59 & 5.98 \\
			Portfolio C -- Consumer & Independent & 3.92 & 1.88 & 0.44 & 3.16 \\
			& Correlated  & 3.93 & 3.53 & 1.65 & 6.22 \\
			Portfolio C -- DevTools & Independent & 4.48 & 1.99 & 0.40 & 3.13 \\
			& Correlated  & 4.50 & 3.71 & 1.57 & 5.96 \\
			Portfolio C -- Health & Independent & 3.49 & 1.78 & 0.47 & 3.19 \\
			& Correlated  & 3.49 & 3.27 & 1.76 & 6.83 \\
			\bottomrule
		\end{tabular}
	\end{adjustbox}
\end{table}

\begin{table}[H]
	\centering
	\caption{Portfolio-Level Tail Probabilities, 40-Deal Portfolios}
	\label{tab:tail_40_full}
	\begin{adjustbox}{max width=\textwidth}
		\begin{tabular}{l l c c c c c}
			\toprule
			Portfolio & Setting & $P(K\ge1)$ & $P(K\ge2)$ & $P(K\ge3)$ & $P(K\ge10)$ & $P(K\ge20)$ \\
			\midrule
			Portfolio A & Independent & 99.63\% & 97.61\% & 91.28\% & 2.97\% & 0.00\% \\
			& Correlated  & 95.22\% & 84.54\% & 70.97\% & 15.34\% & 0.64\% \\
			Portfolio B & Independent & 99.98\% & 99.64\% & 98.23\% & 16.18\% & 0.00\% \\
			& Correlated  & 98.61\% & 94.23\% & 86.47\% & 25.10\% & 2.38\% \\
			Portfolio C (Diversified) & Independent & 99.08\% & 94.63\% & 83.63\% & 1.04\% & 0.00\% \\
			& Correlated  & 93.52\% & 79.94\% & 64.05\% & 10.25\% & 0.32\% \\
			Portfolio C -- SaaS & Independent & 99.27\% & 95.46\% & 85.41\% & 1.34\% & 0.00\% \\
			& Correlated  & 94.12\% & 81.35\% & 65.41\% & 11.37\% & 0.37\% \\
			Portfolio C -- AI & Independent & 98.92\% & 93.93\% & 81.87\% & 0.81\% & 0.00\% \\
			& Correlated  & 92.52\% & 78.03\% & 61.29\% & 10.08\% & 0.31\% \\
			Portfolio C -- Fintech & Independent & 98.82\% & 93.50\% & 80.91\% & 0.75\% & 0.00\% \\
			& Correlated  & 92.87\% & 78.58\% & 61.88\% & 9.29\% & 0.30\% \\
			Portfolio C -- Consumer & Independent & 98.36\% & 91.52\% & 76.86\% & 0.48\% & 0.00\% \\
			& Correlated  & 91.07\% & 74.83\% & 57.53\% & 8.12\% & 0.23\% \\
			Portfolio C -- DevTools & Independent & 99.14\% & 94.96\% & 84.10\% & 1.12\% & 0.00\% \\
			& Correlated  & 94.29\% & 81.86\% & 65.85\% & 9.89\% & 0.41\% \\
			Portfolio C -- Health & Independent & 97.47\% & 87.63\% & 68.93\% & 0.20\% & 0.00\% \\
			& Correlated  & 88.95\% & 70.36\% & 51.78\% & 6.32\% & 0.13\% \\
			\bottomrule
		\end{tabular}
	\end{adjustbox}
\end{table}

The distributional analysis above captures the full portfolio construction insight of the paper; translating these outcome distributions into monetary returns would require deal-level return assumptions that are beyond the scope of this framework and left for future work.

\section{Conclusion}

This study develops a unified Gaussian copula framework to model deal-level dependence in VC portfolios and to quantify its implications for aggregate outcomes. By learning correlation structures directly from observed joint realizations, the framework provides a coherent explanation for the heavy-tailed nature of venture returns, even when marginal success probabilities remain moderate. The analysis demonstrates that dependence plays a first-order role in shaping portfolio behavior. Correlation does not materially alter expected returns, but it substantially increases dispersion, right-tail mass, and higher-moment risk.

Extreme portfolio outcomes may arise from correlated realizations rather than purely from superior average deal quality. While diversification across founders, geographies, and markets attenuates tail amplification, it does not eliminate it. Conversely, concentrated portfolios exhibit stronger upper-tail potential, such as the portfolio with all repeat founders in California, but simultaneously experience greater volatility and a higher probability of disappointing aggregate outcomes. These findings underscore that effective venture portfolio construction requires explicit management of dependence structures, rather than an exclusive focus on improving marginal deal selection. In environments characterized by systemic shocks and shared risk factors, correlation can be a key driver of both exceptional success and pronounced downside risk.

This study has several limitations. First, the Gaussian copula has an upper tail dependence coefficient of exactly zero for any finite latent correlation. This means that in the limit, extreme joint successes are no more probable than under independence, which may understate the very tail events most relevant for top-decile VC fund performance. The t-copula or Gumbel copula family, which exhibit non-zero upper tail dependence, are natural extensions. Second, the estimated global factor parameter $\hat{\alpha}_0 = 0$ reflects a boundary solution that is partly an identification issue rather than a substantive empirical finding, as discussed in Section 4.3; macro-level common shocks such as vintage effects and interest rate cycles are not captured by the current attribute-level framework. Third, the framework is static and does not incorporate time dynamics or vintage effects, which almost certainly cause the dependence structure to vary across investment cycles. Fourth, the mutual exclusivity of founder type and geography indicators introduces sum-to-one constraints that are not accounted for in estimation; dropping one reference category per exclusive group would yield a more cleanly identified $\Sigma$. Finally, deal-level success probabilities are constructed synthetically from a selected dataset with observed success rates approximately five times higher than the population-level base rate; the core qualitative findings are robust to this gap, as argued in Section 4.2, but absolute simulation quantities should not be interpreted as population-representative.

Future research could extend the framework in several directions. Incorporating time dynamics and vintage effects would allow dependence between deals to evolve across investment cycles. Improving the estimation of deal-level success probabilities using richer data could enhance the realism of the joint probability structure, while expanded stress testing across different market environments could strengthen robustness. Extending the framework to include features such as multi-round investing, capital pacing, and follow-on decisions would further improve its applicability for venture capital portfolio analysis.

\bibliographystyle{apalike}
\bibliography{references}

\newpage
\appendix

\section*{Appendix}
\addcontentsline{toc}{section}{Appendix}

 \section{Additional Portfolio Distribution Results}
 \label{app:dist}
 
 \begin{figure}[H]
 	\centering
 	\begin{subfigure}{0.32\textwidth}
 		\includegraphics[width=\textwidth]{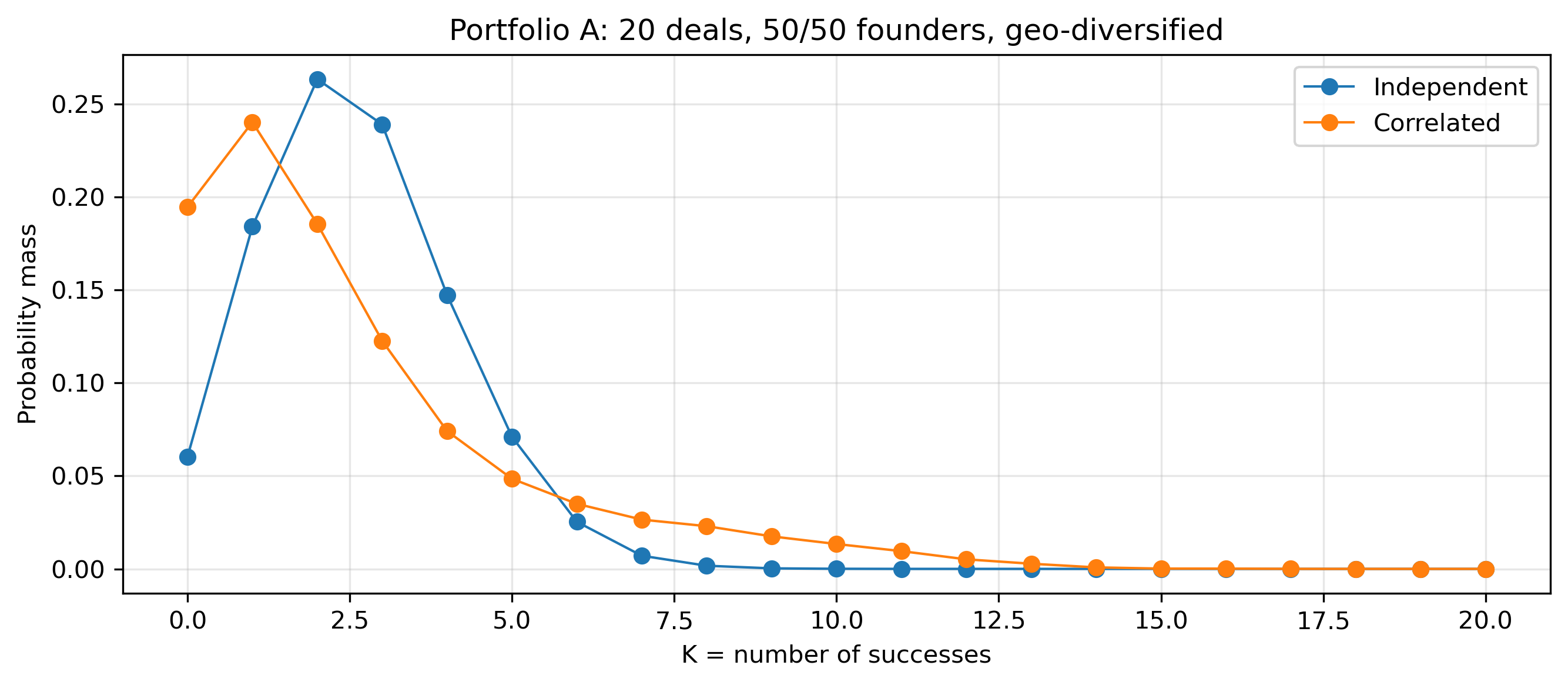}
 		\caption{20 deals}
 	\end{subfigure}
 	\hfill
 	\begin{subfigure}{0.32\textwidth}
 		\includegraphics[width=\textwidth]{graphs/Portfolio_A__40_deals__50_50_founders__geo-diversified.png}
 		\caption{40 deals}
 	\end{subfigure}
 	\hfill
 	\begin{subfigure}{0.32\textwidth}
 		\includegraphics[width=\textwidth]{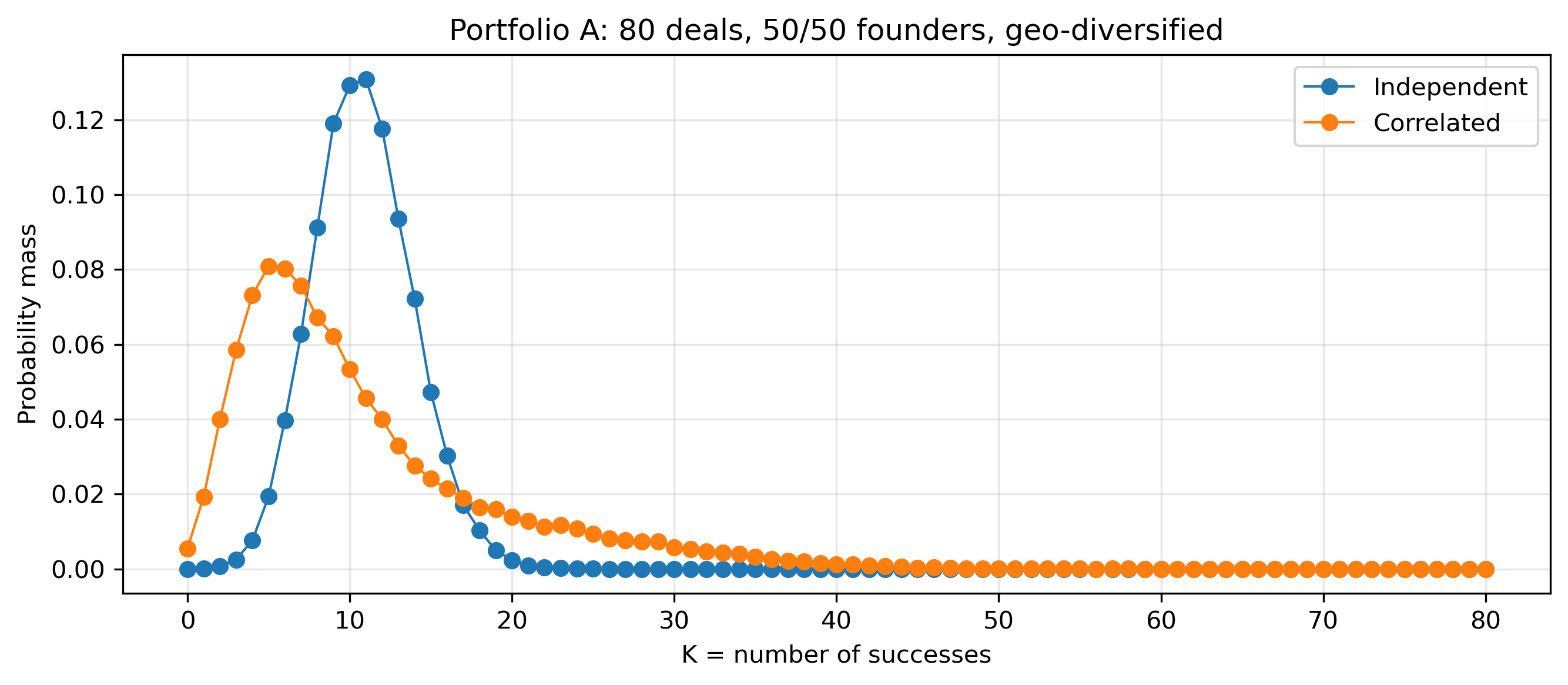}
 		\caption{80 deals}
 	\end{subfigure}
 	\caption{Distribution of the number of successful deals for Portfolio A (50/50 founders, geographically diversified).}
 	\label{fig:portfolio_A}
 \end{figure}
 
 \begin{figure}[H]
 	\centering
 	\begin{subfigure}{0.32\textwidth}
 		\includegraphics[width=\textwidth]{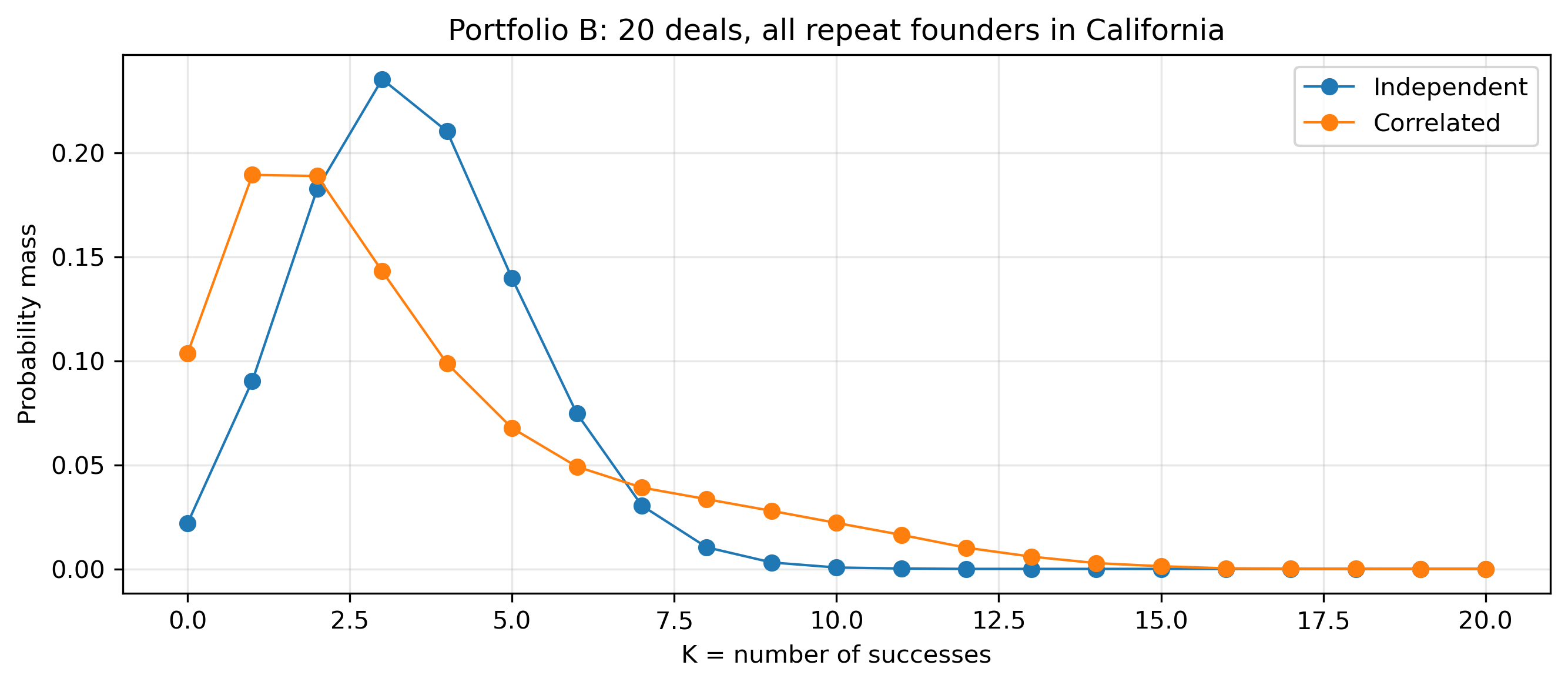}
 		\caption{20 deals}
 	\end{subfigure}
 	\hfill
 	\begin{subfigure}{0.32\textwidth}
 		\includegraphics[width=\textwidth]{graphs/Portfolio_B__40_deals__all_repeat_founders_in_California.png}
 		\caption{40 deals}
 	\end{subfigure}
 	\hfill
 	\begin{subfigure}{0.32\textwidth}
 		\includegraphics[width=\textwidth]{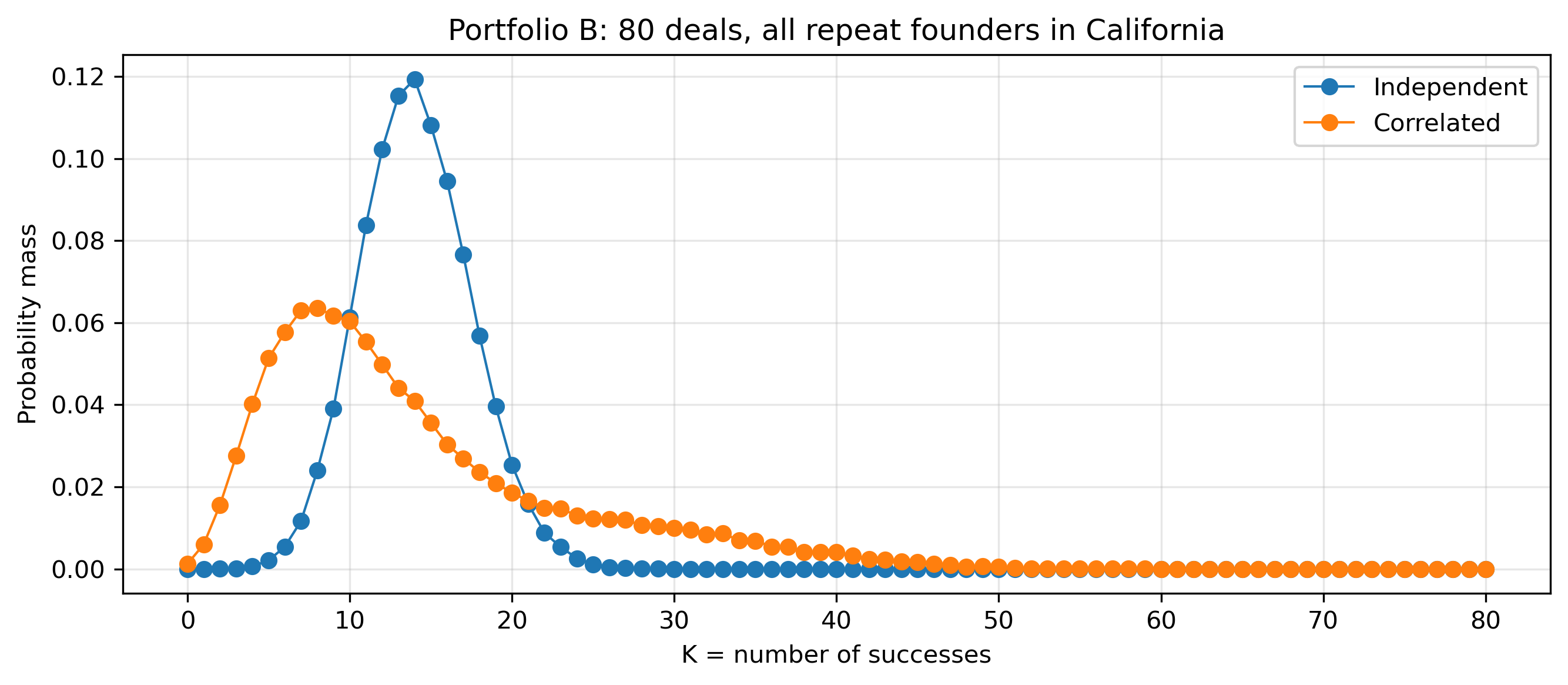}
 		\caption{80 deals}
 	\end{subfigure}
 	\caption{Distribution of the number of successful deals for Portfolio B (all repeat founders in California).}
 	\label{fig:portfolio_B}
 \end{figure}
 
 \begin{figure}[H]
 	\centering
 	\begin{subfigure}{0.32\textwidth}
 		\includegraphics[width=\textwidth]{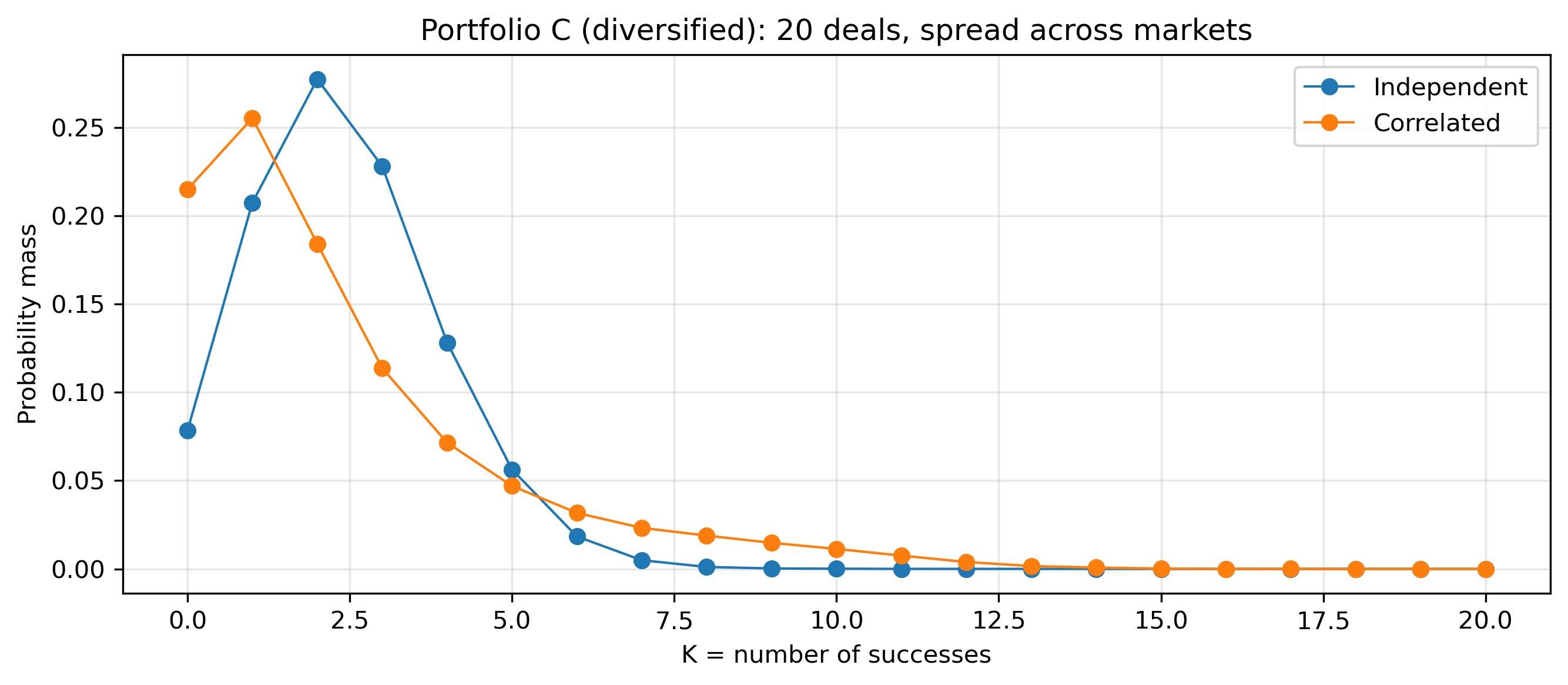}
 		\caption{20 deals}
 	\end{subfigure}
 	\hfill
 	\begin{subfigure}{0.32\textwidth}
 		\includegraphics[width=\textwidth]{graphs/Portfolio_C__diversified___40_deals__spread_across_markets.png}
 		\caption{40 deals}
 	\end{subfigure}
 	\hfill
 	\begin{subfigure}{0.32\textwidth}
 		\includegraphics[width=\textwidth]{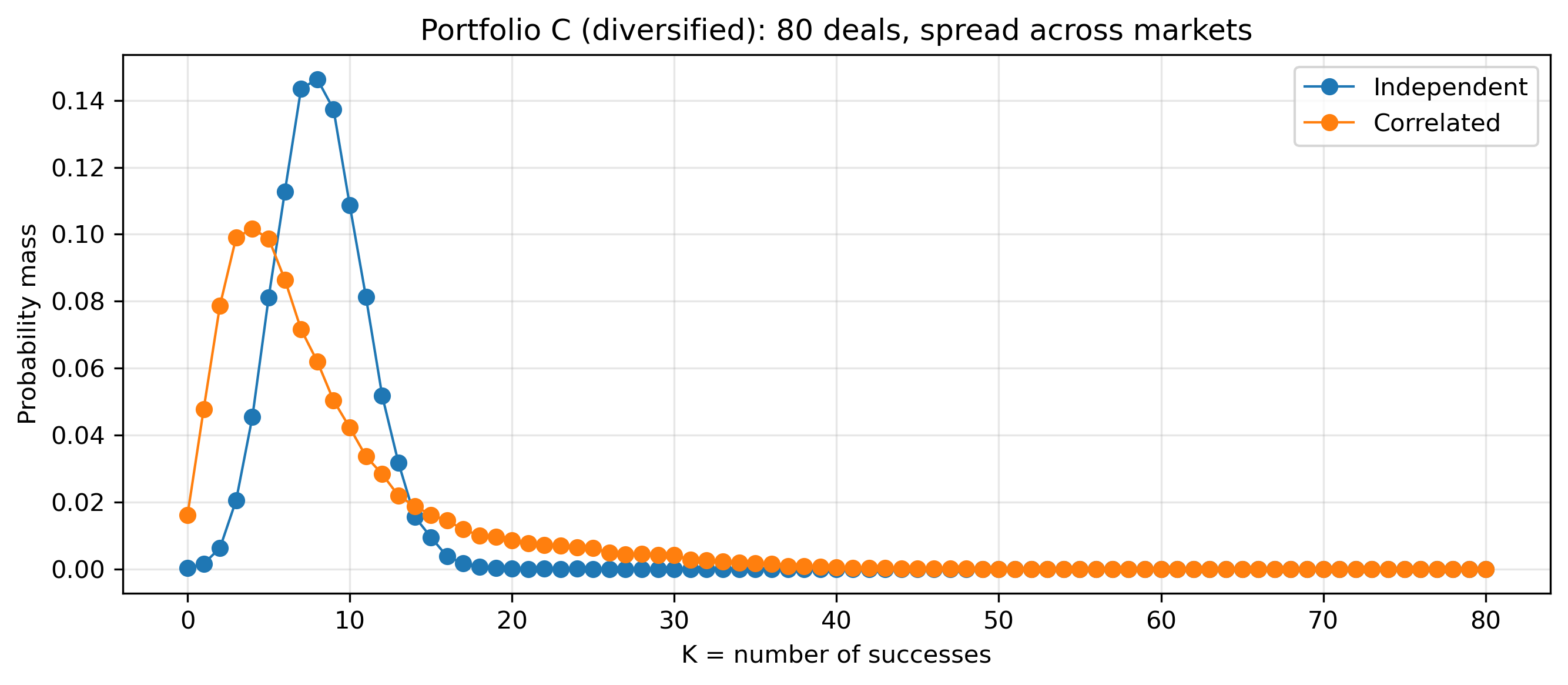}
 		\caption{80 deals}
 	\end{subfigure}
 	\caption{Distribution of the number of successful deals for diversified Portfolio C. Market diversification mitigates but does not eliminate correlation-driven tail amplification.}
 	\label{fig:portfolio_C_diversified}
 \end{figure}

 \begin{figure}[H]
 	\centering
 	\begin{subfigure}{0.32\textwidth}
 		\includegraphics[width=\textwidth]{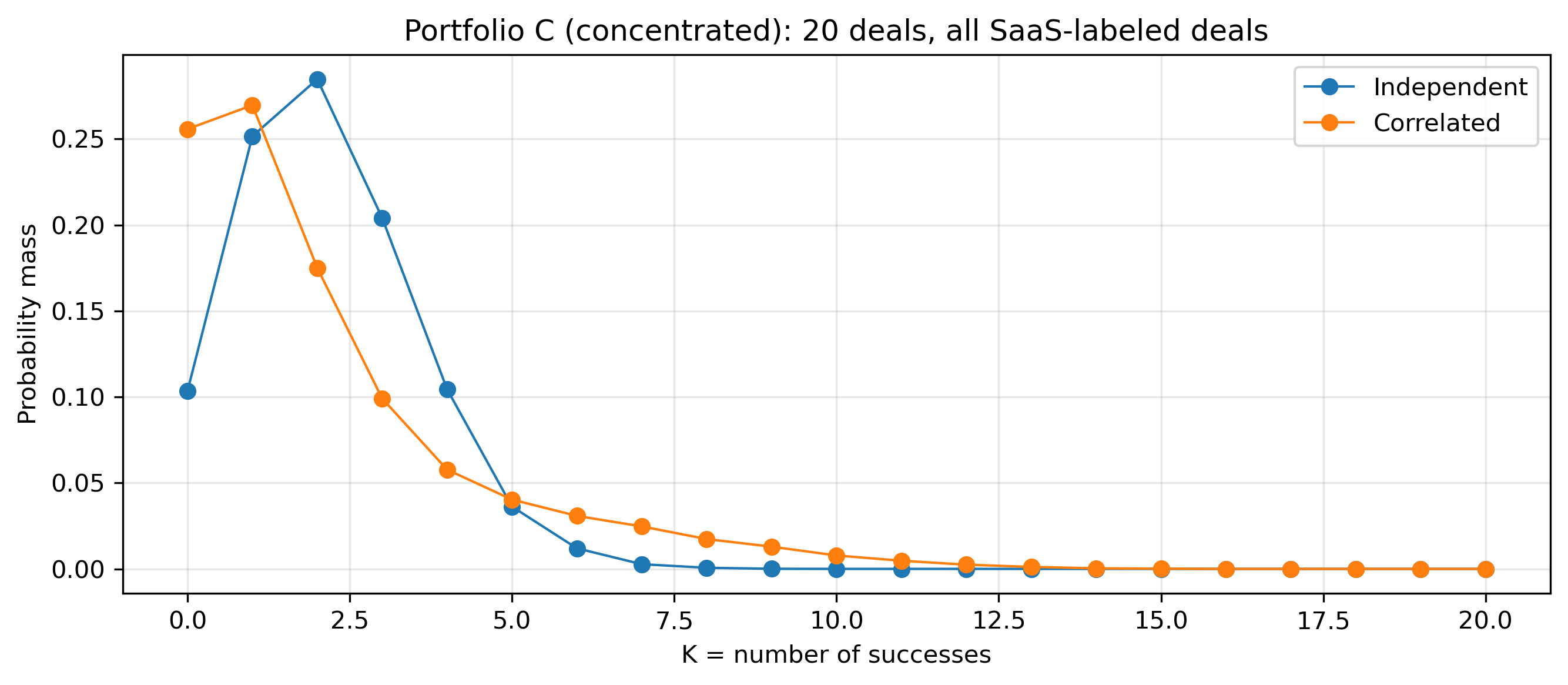}
 		\caption{SaaS}
 	\end{subfigure}
 	\hfill
 	\begin{subfigure}{0.32\textwidth}
 		\includegraphics[width=\textwidth]{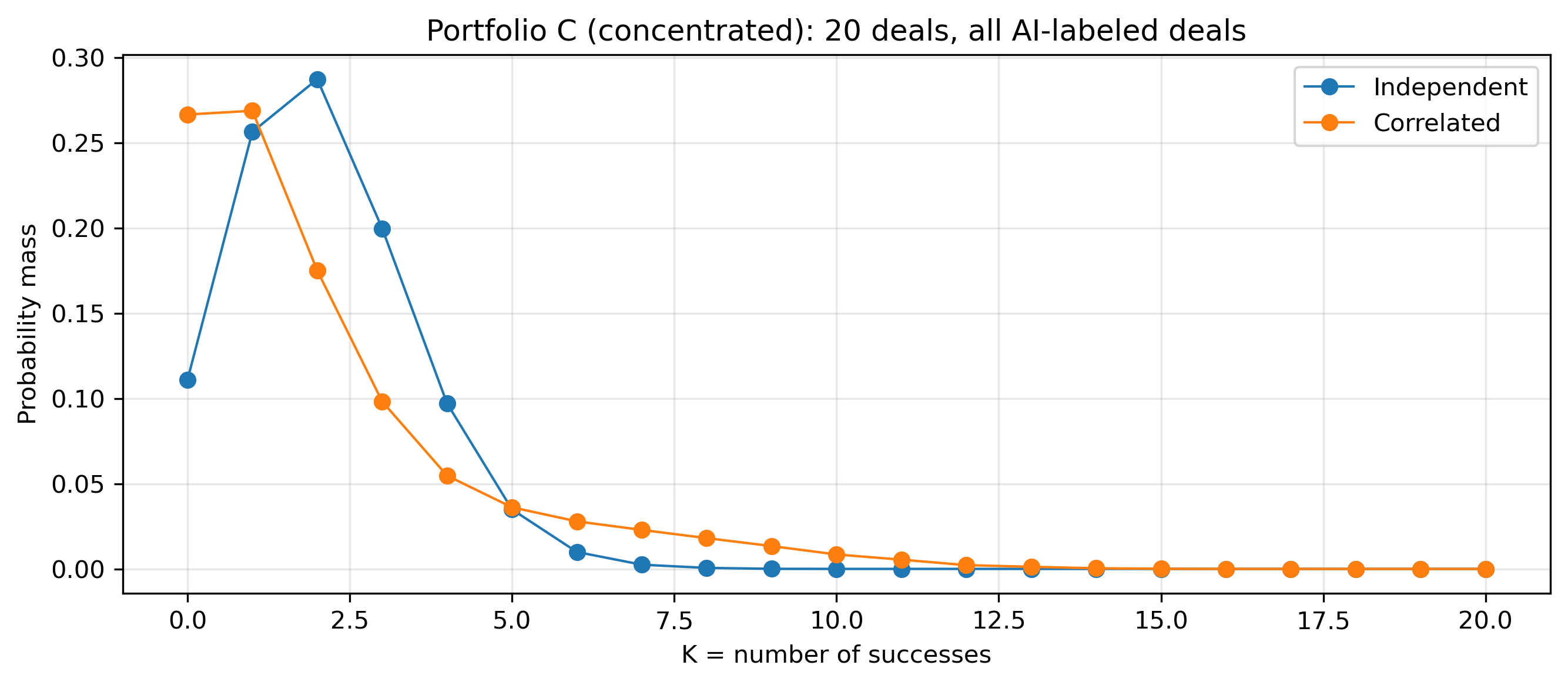}
 		\caption{AI}
 	\end{subfigure}
 	\hfill
 	\begin{subfigure}{0.32\textwidth}
 		\includegraphics[width=\textwidth]{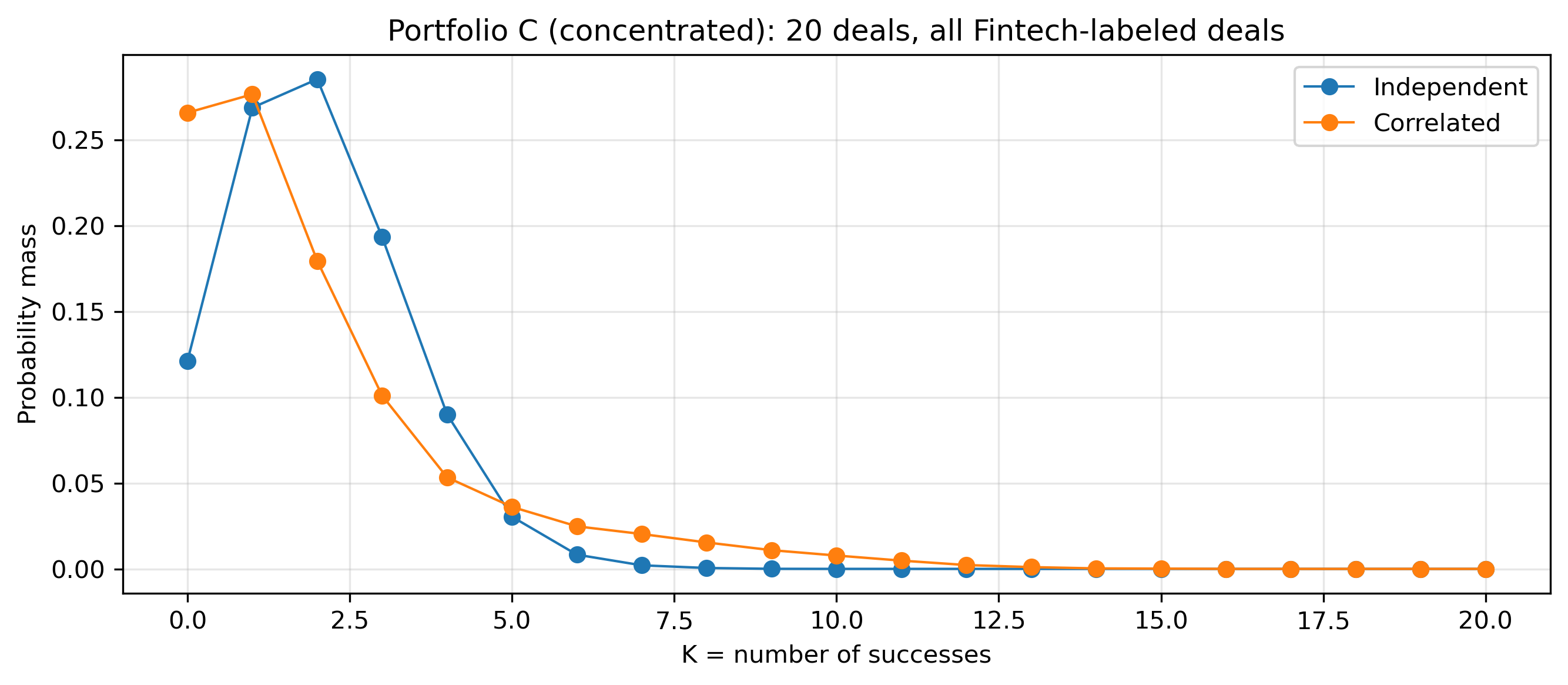}
 		\caption{Fintech}
 	\end{subfigure}
 	
 	\vspace{0.5em}
 	
 	\begin{subfigure}{0.32\textwidth}
 		\includegraphics[width=\textwidth]{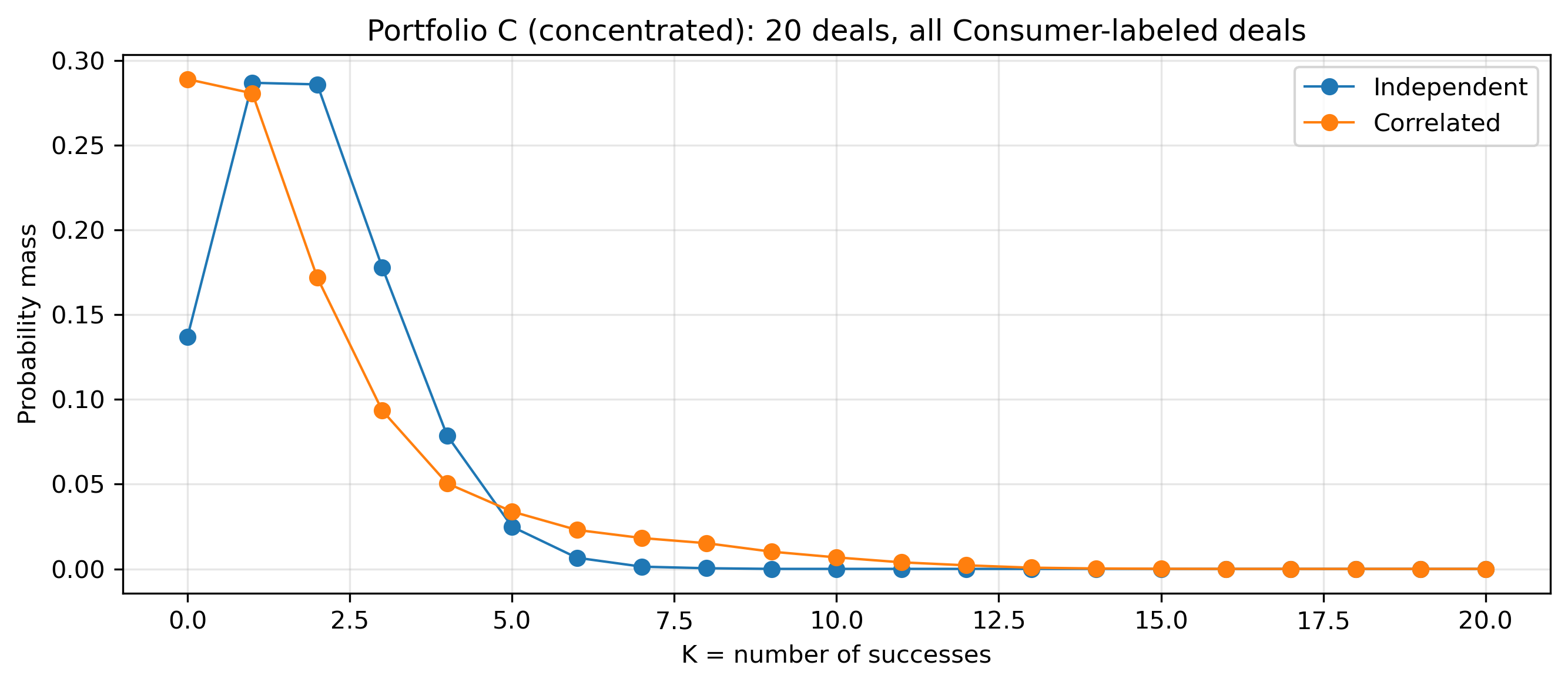}
 		\caption{Consumer}
 	\end{subfigure}
 	\hfill
 	\begin{subfigure}{0.32\textwidth}
 		\includegraphics[width=\textwidth]{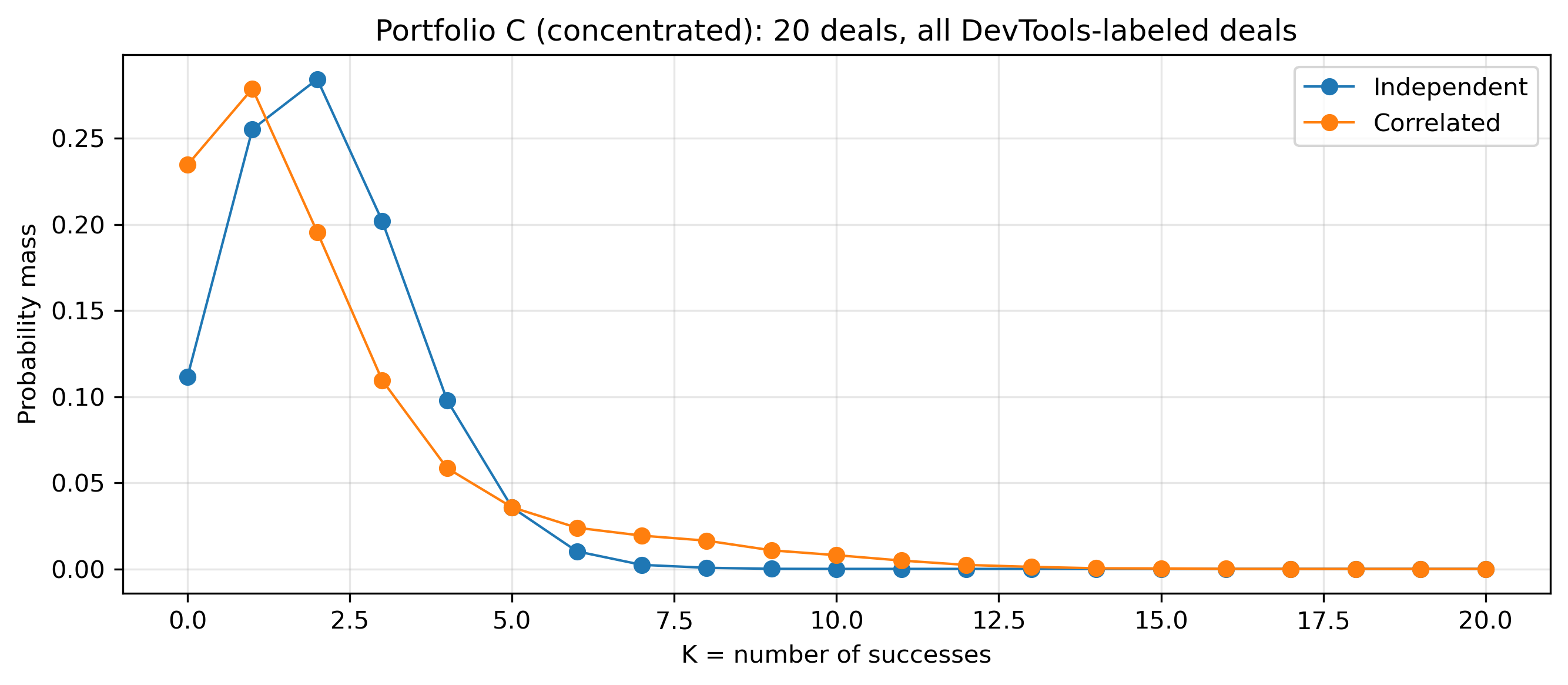}
 		\caption{DevTools}
 	\end{subfigure}
 	\hfill
 	\begin{subfigure}{0.32\textwidth}
 		\includegraphics[width=\textwidth]{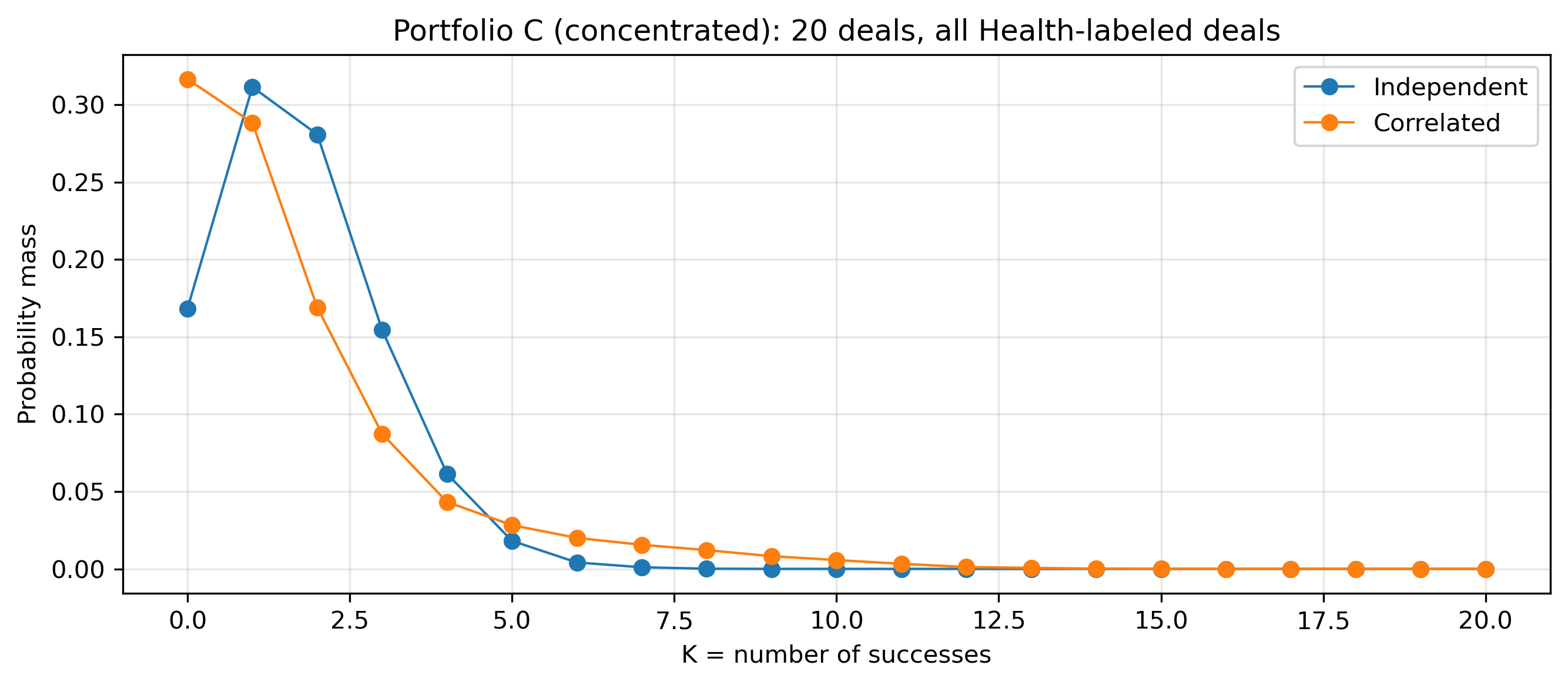}
 		\caption{Health}
 	\end{subfigure}
 	\caption{Concentrated Portfolio C with 20 deals across market categories.}
 	\label{fig:portfolio_C_20_appendix}
 \end{figure}
 
 \begin{figure}[H]
 	\centering
 	\begin{subfigure}{0.32\textwidth}
 		\includegraphics[width=\textwidth]{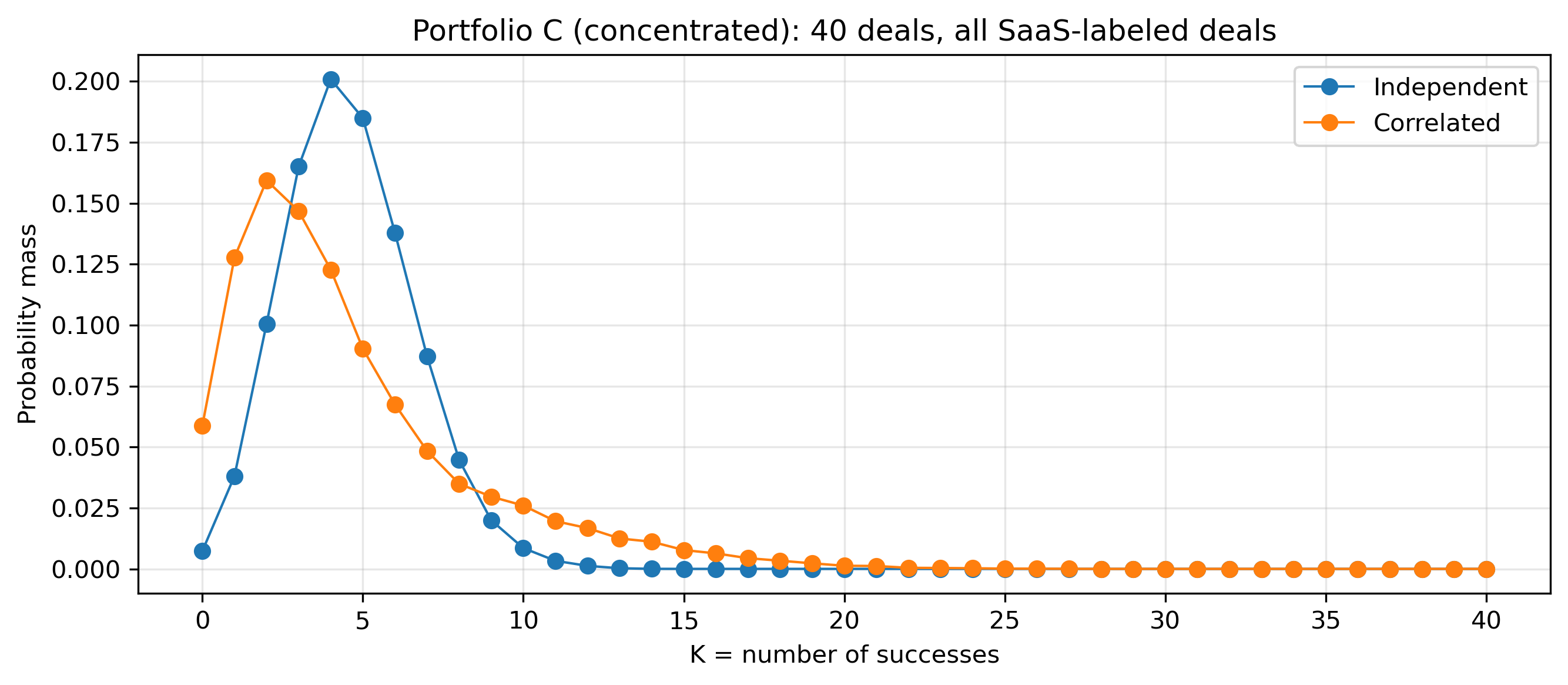}
 		\caption{SaaS}
 	\end{subfigure}
 	\hfill
 	\begin{subfigure}{0.32\textwidth}
 		\includegraphics[width=\textwidth]{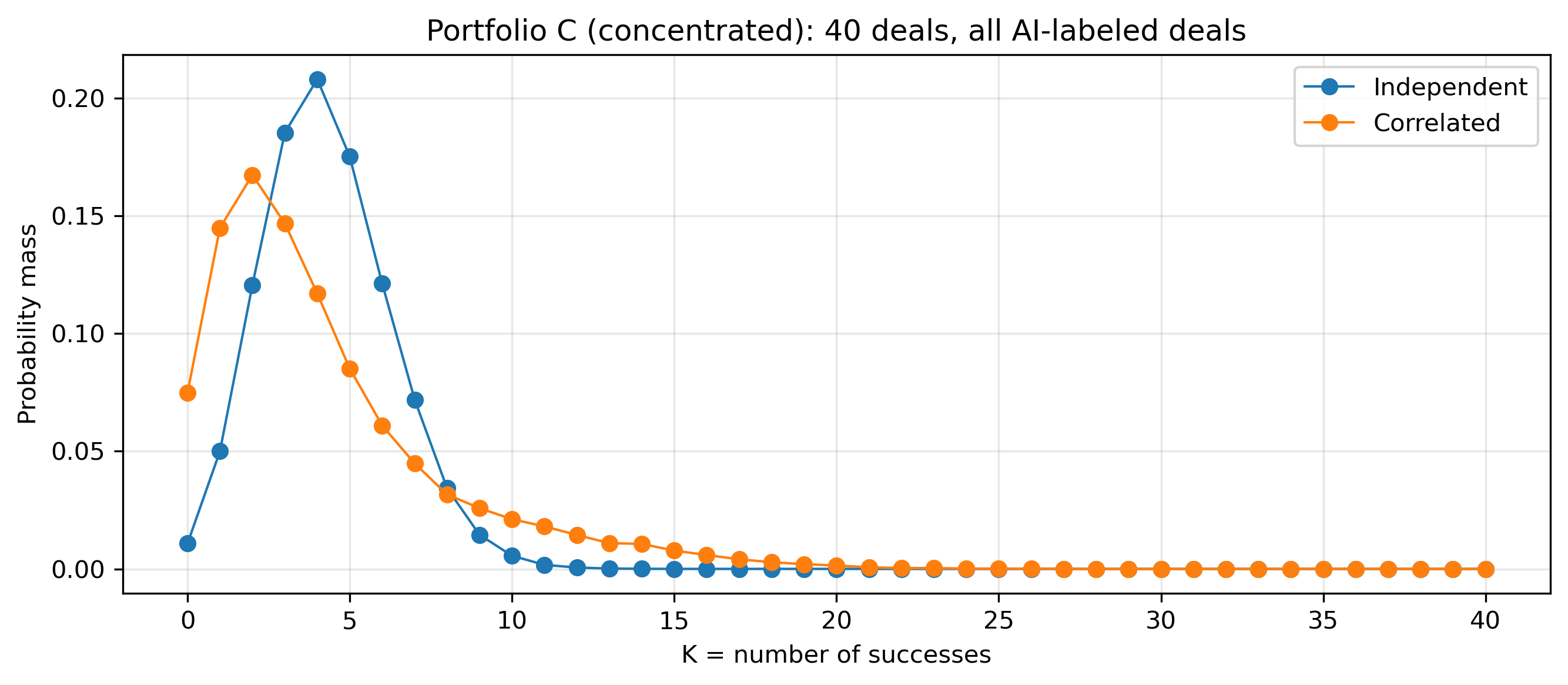}
 		\caption{AI}
 	\end{subfigure}
 	\hfill
 	\begin{subfigure}{0.32\textwidth}
 		\includegraphics[width=\textwidth]{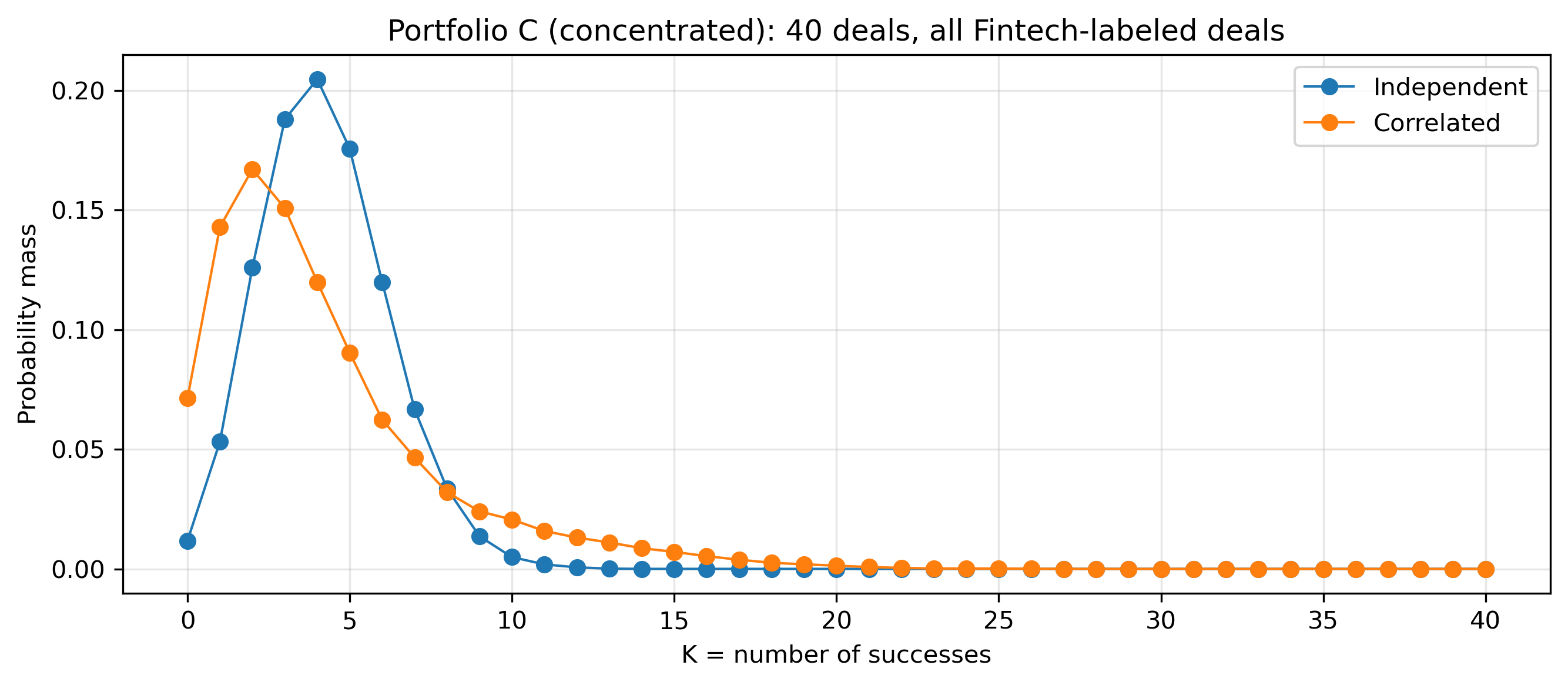}
 		\caption{Fintech}
 	\end{subfigure}
 	
 	\vspace{0.5em}
 	
 	\begin{subfigure}{0.32\textwidth}
 		\includegraphics[width=\textwidth]{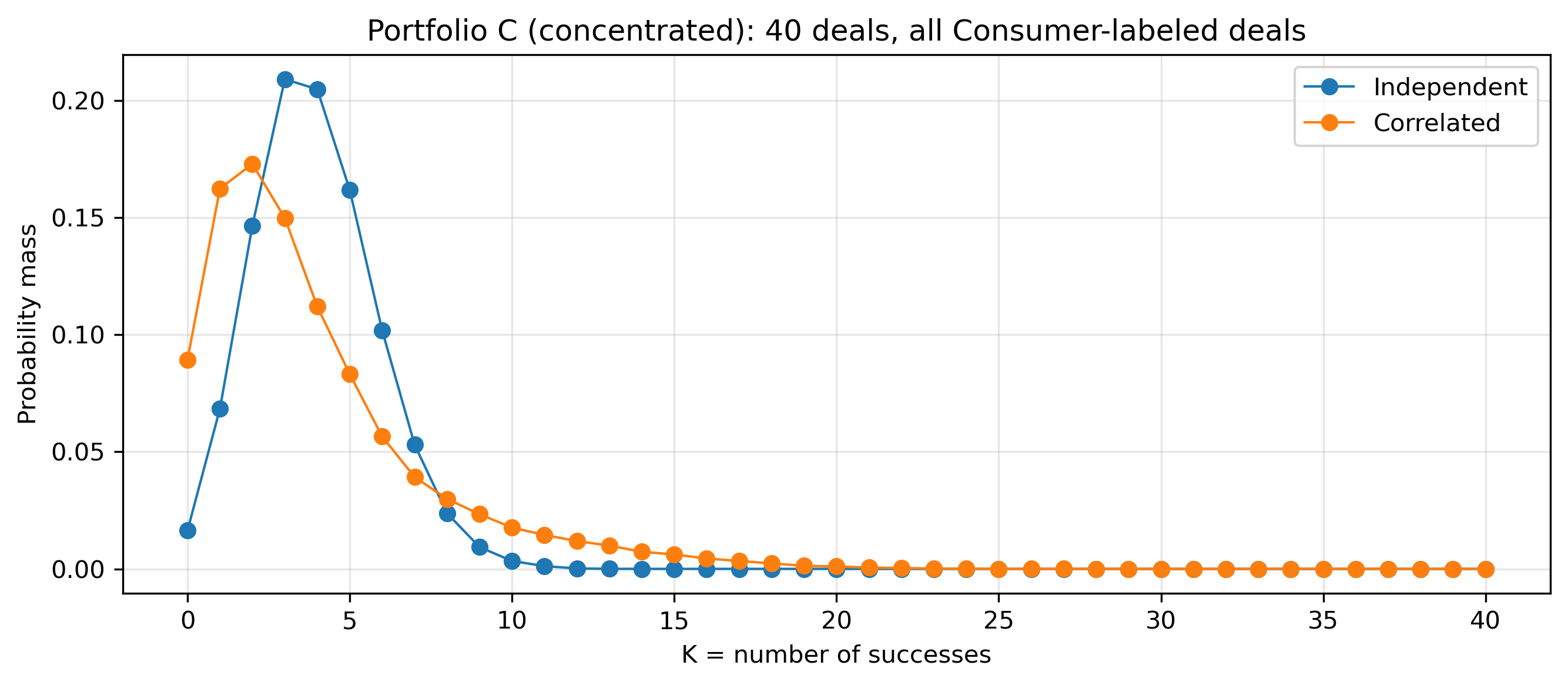}
 		\caption{Consumer}
 	\end{subfigure}
 	\hfill
 	\begin{subfigure}{0.32\textwidth}
 		\includegraphics[width=\textwidth]{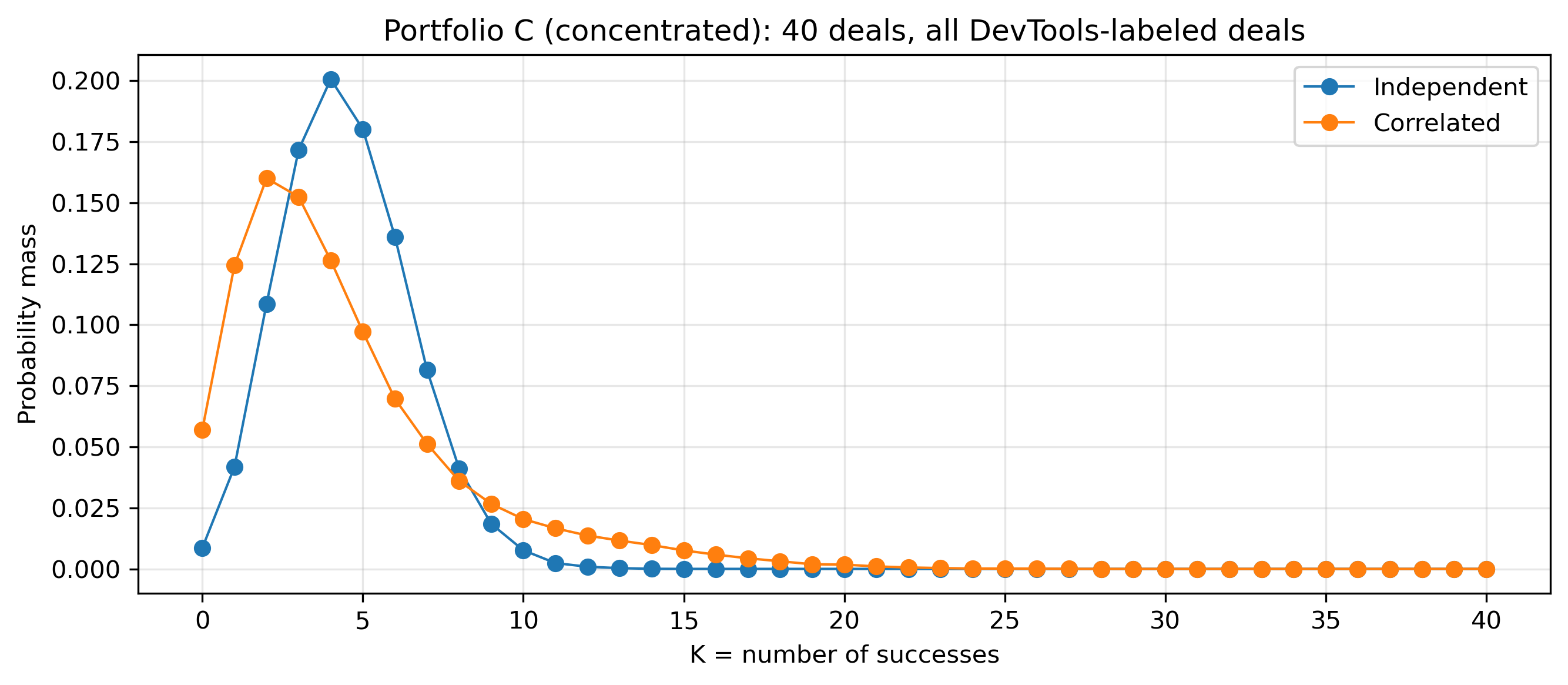}
 		\caption{DevTools}
 	\end{subfigure}
 	\hfill
 	\begin{subfigure}{0.32\textwidth}
 		\includegraphics[width=\textwidth]{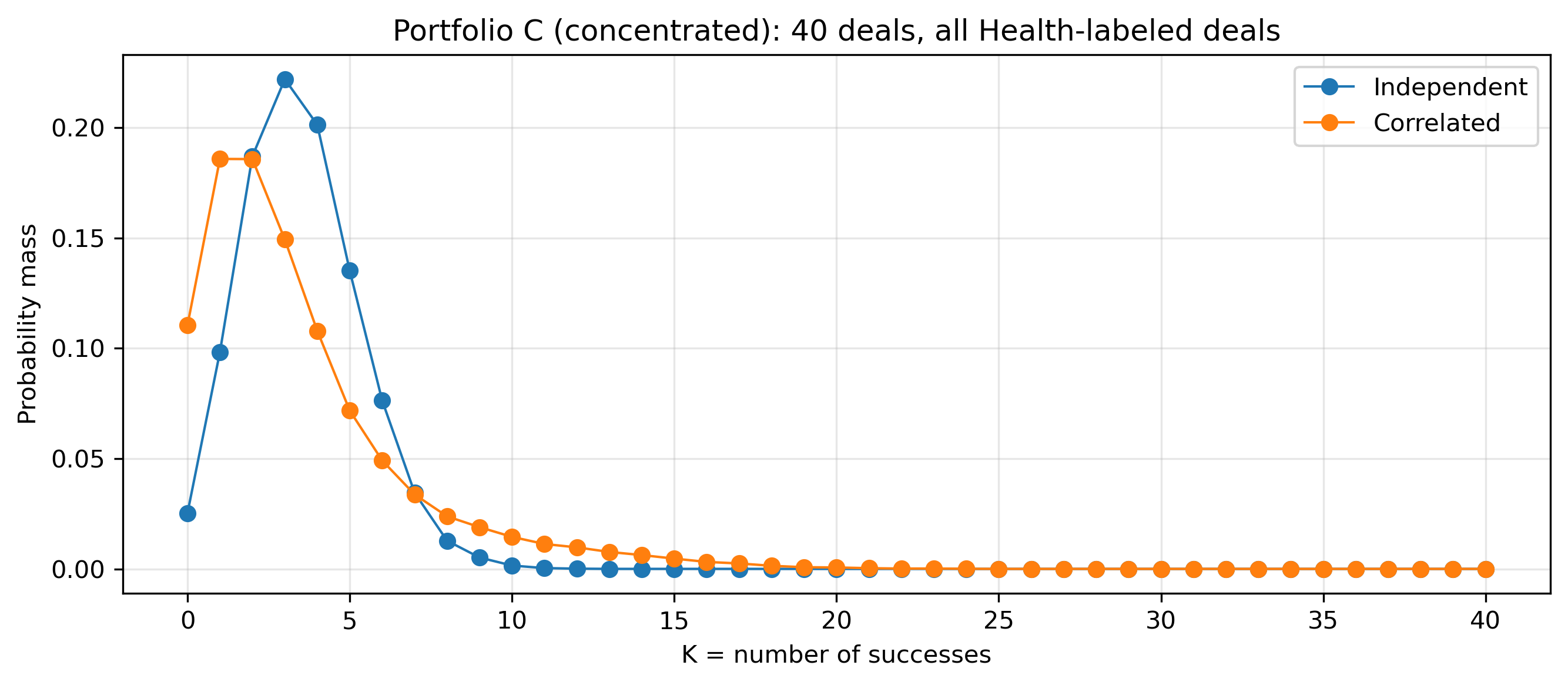}
 		\caption{Health}
 	\end{subfigure}
 	\caption{Concentrated Portfolio C with 40 deals across market categories.}
 	\label{fig:portfolio_C_40_appendix}
 \end{figure}
 
 \begin{figure}[H]
 	\centering
 	\begin{subfigure}{0.32\textwidth}
 		\includegraphics[width=\textwidth]{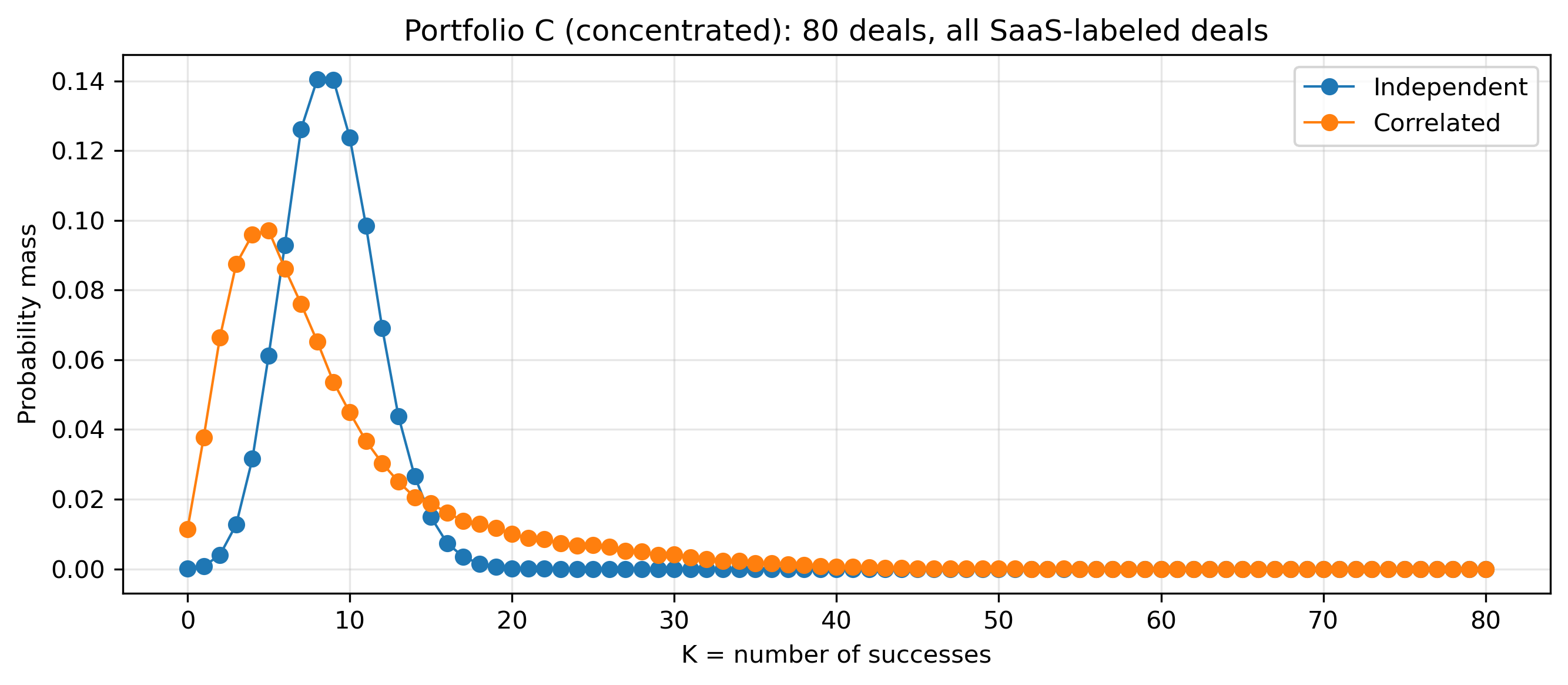}
 		\caption{SaaS}
 	\end{subfigure}
 	\hfill
 	\begin{subfigure}{0.32\textwidth}
 		\includegraphics[width=\textwidth]{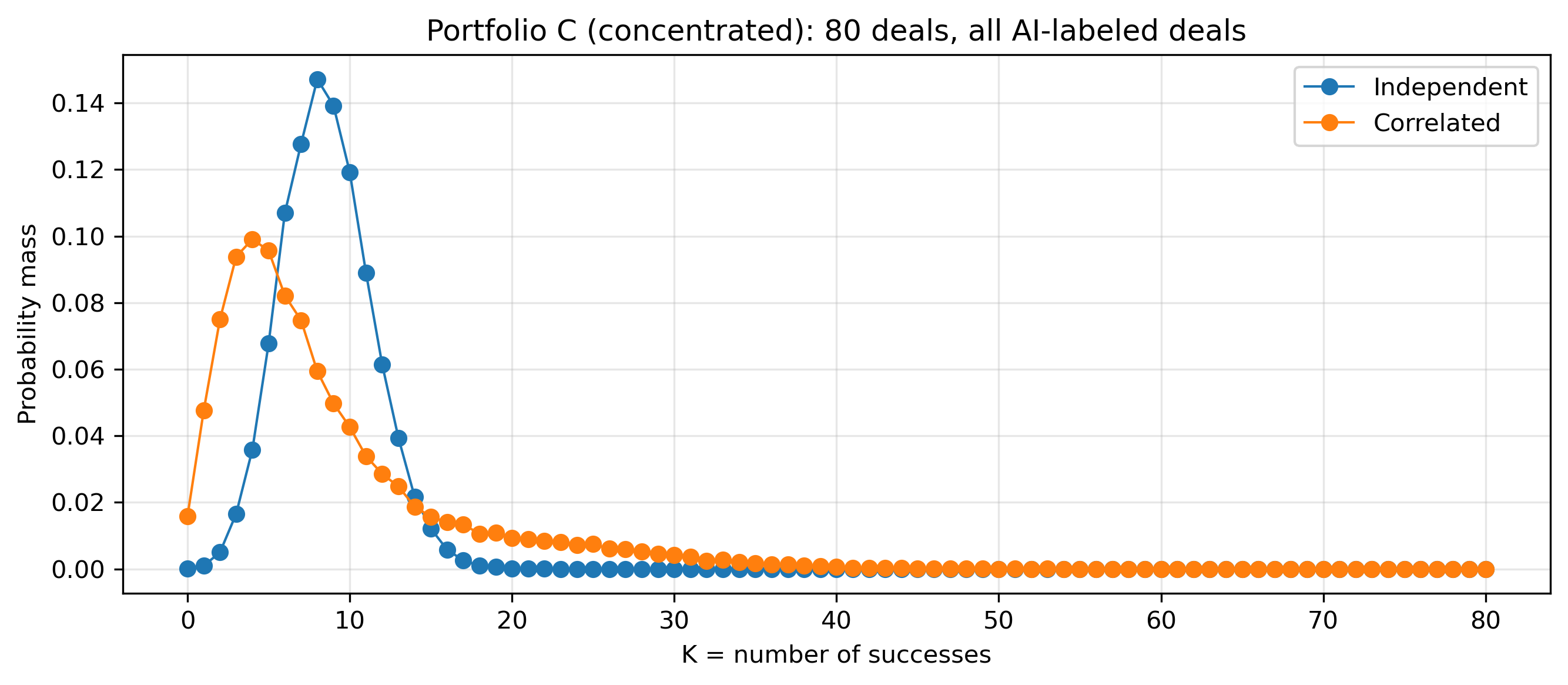}
 		\caption{AI}
 	\end{subfigure}
 	\hfill
 	\begin{subfigure}{0.32\textwidth}
 		\includegraphics[width=\textwidth]{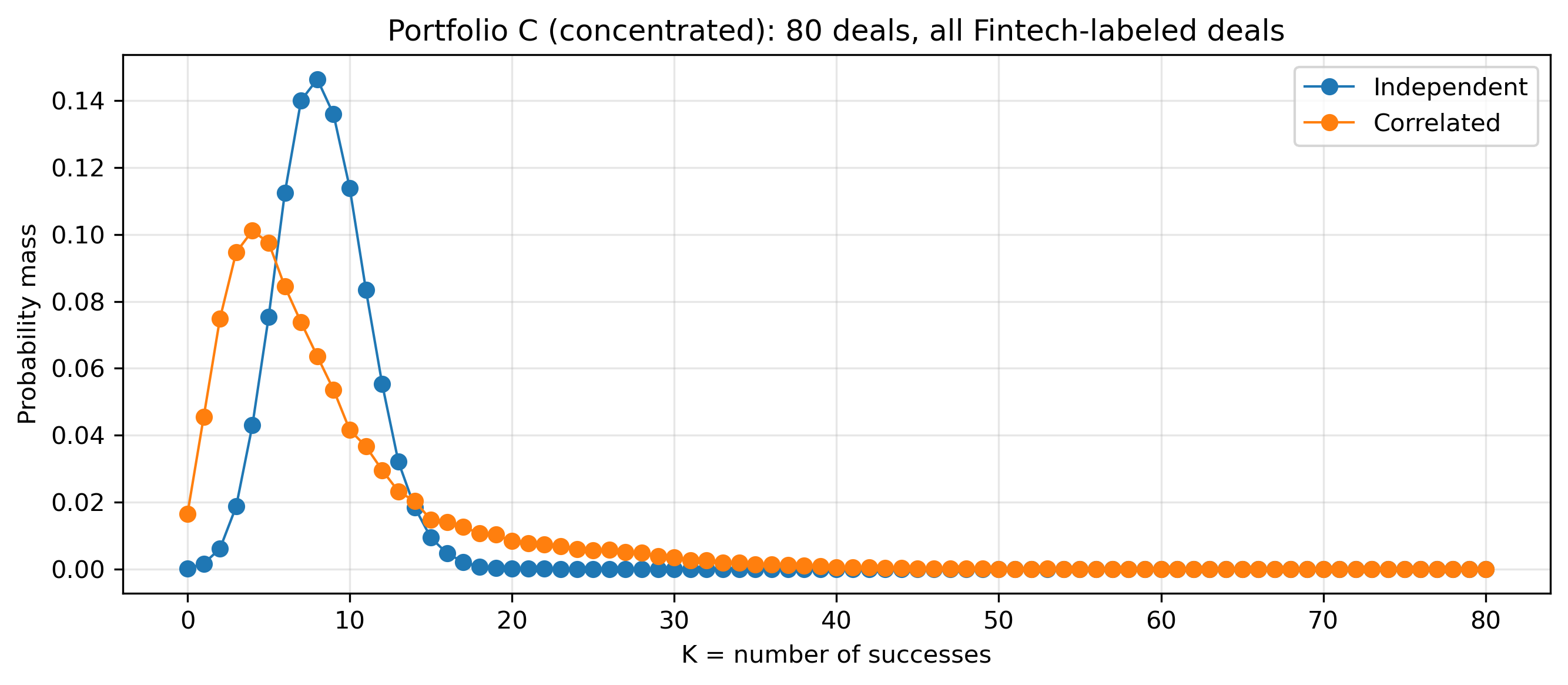}
 		\caption{Fintech}
 	\end{subfigure}
 	
 	\vspace{0.5em}
 	
 	\begin{subfigure}{0.32\textwidth}
 		\includegraphics[width=\textwidth]{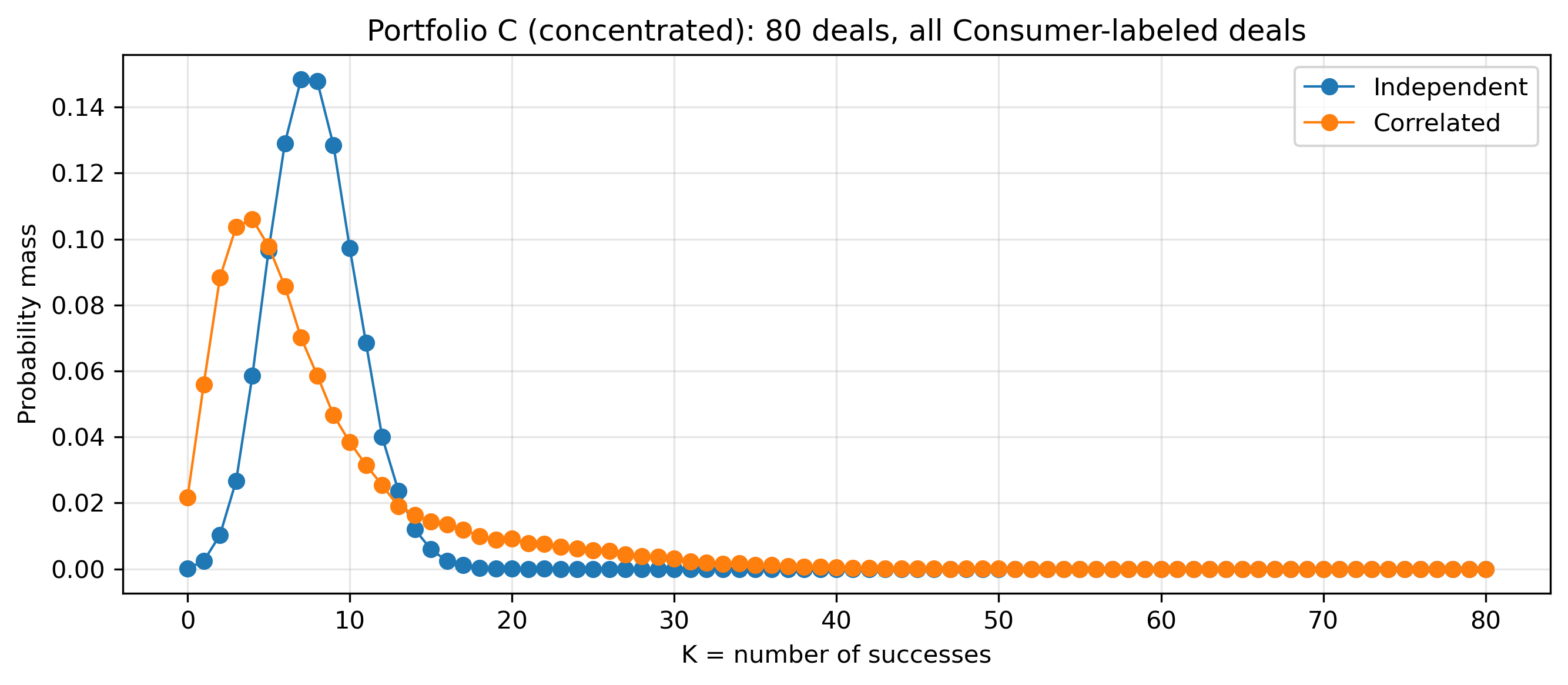}
 		\caption{Consumer}
 	\end{subfigure}
 	\hfill
 	\begin{subfigure}{0.32\textwidth}
 		\includegraphics[width=\textwidth]{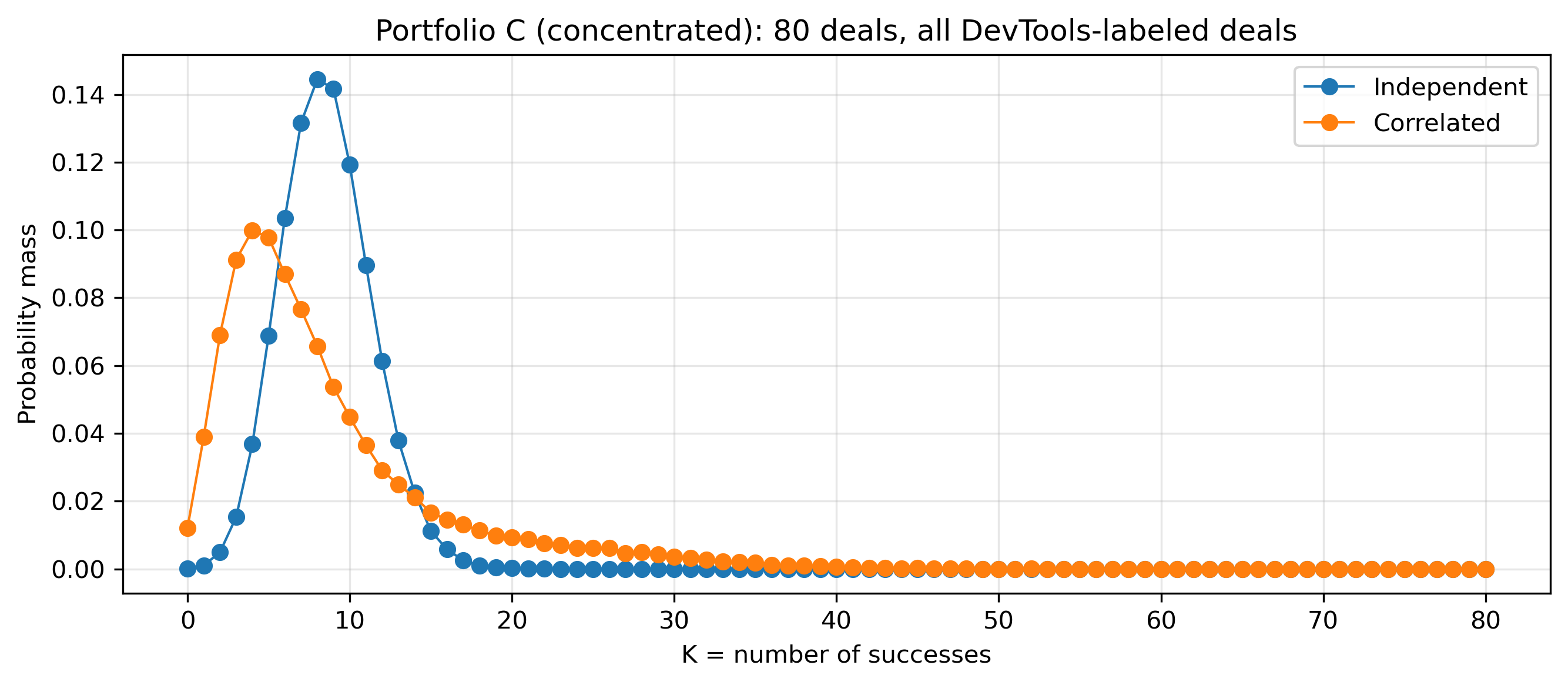}
 		\caption{DevTools}
 	\end{subfigure}
 	\hfill
 	\begin{subfigure}{0.32\textwidth}
 		\includegraphics[width=\textwidth]{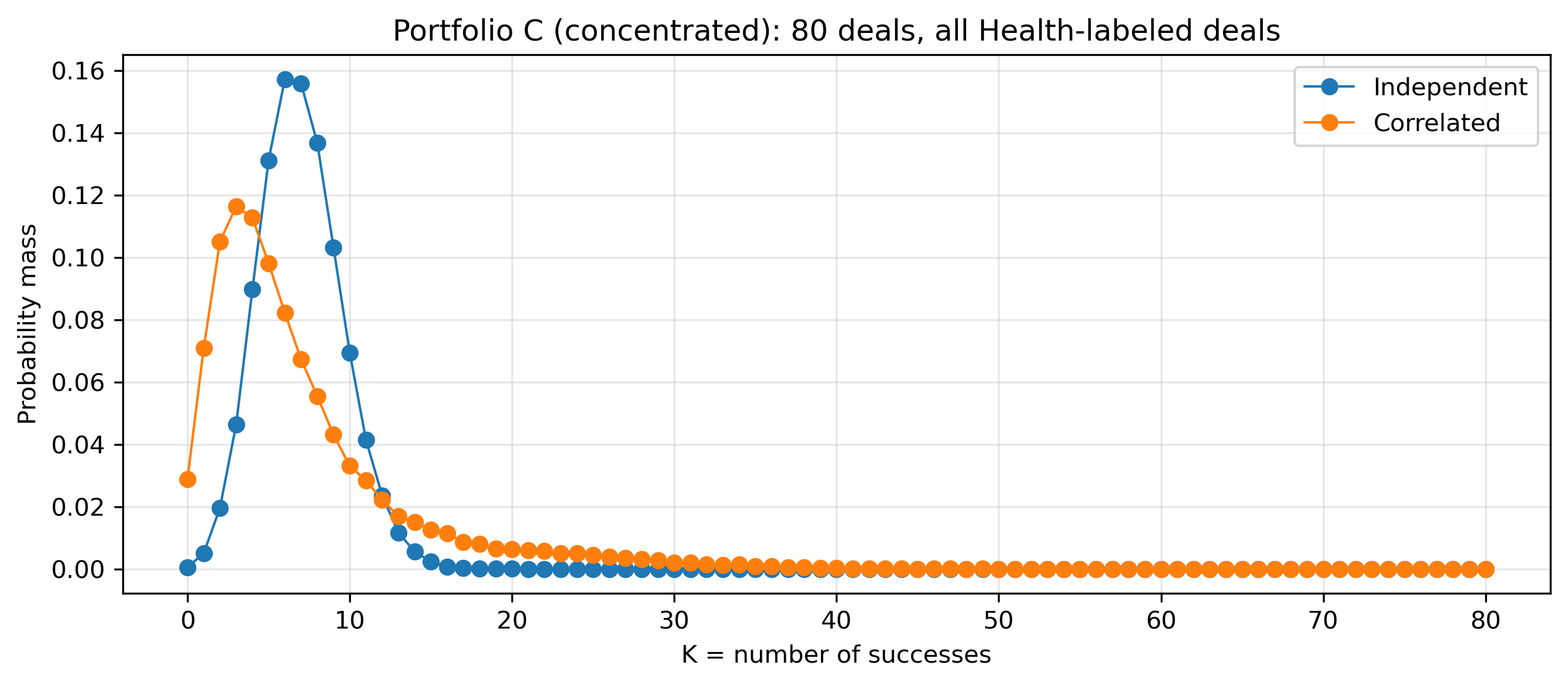}
 		\caption{Health}
 	\end{subfigure}
 	\caption{Concentrated Portfolio C with 80 deals across market categories.}
 	\label{fig:portfolio_C_80_appendix}
 \end{figure}

 \begin{table}[H]
 	\centering
 	\caption{Summary Statistics of the Number of Successful Deals ($K$), 20-Deal Portfolios}
 	\label{tab:K_summary_20_appendix}
 	\begin{adjustbox}{max width=\textwidth}
 		\begin{tabular}{l l c c c c}
 			\toprule
 			Portfolio & Setting & Mean & Std & Skew & Kurt \\
 			\midrule
 			Portfolio A & Independent & 2.59 & 1.48 & 0.47 & 3.11 \\
 			& Correlated  & 2.61 & 2.66 & 1.51 & 5.16 \\
 			Portfolio B & Independent & 3.48 & 1.69 & 0.38 & 3.07 \\
 			& Correlated  & 3.48 & 3.01 & 1.21 & 4.06 \\
 			Portfolio C (Diversified) & Independent & 2.40 & 1.44 & 0.52 & 3.22 \\
 			& Correlated  & 2.40 & 2.52 & 1.60 & 5.62 \\
 			Portfolio C -- SaaS & Independent & 2.13 & 1.37 & 0.57 & 3.25 \\
 			& Correlated  & 2.14 & 2.38 & 1.66 & 5.82 \\
 			Portfolio C -- AI & Independent & 2.08 & 1.36 & 0.58 & 3.26 \\
 			& Correlated  & 2.10 & 2.39 & 1.72 & 6.01 \\
 			Portfolio C -- Fintech & Independent & 2.00 & 1.34 & 0.59 & 3.29 \\
 			& Correlated  & 2.02 & 2.30 & 1.81 & 6.57 \\
 			Portfolio C -- Consumer & Independent & 1.88 & 1.30 & 0.62 & 3.30 \\
 			& Correlated  & 1.90 & 2.23 & 1.86 & 6.80 \\
 			Portfolio C -- DevTools & Independent & 2.08 & 1.36 & 0.57 & 3.23 \\
 			& Correlated  & 2.10 & 2.29 & 1.81 & 6.73 \\
 			Portfolio C -- Health & Independent & 1.71 & 1.25 & 0.67 & 3.37 \\
 			& Correlated  & 1.72 & 2.10 & 2.00 & 7.65 \\
 			\bottomrule
 		\end{tabular}
 	\end{adjustbox}
 \end{table}
 
 \begin{table}[H]
 	\centering
 	\caption{Summary Statistics of the Number of Successful Deals ($K$), 40-Deal Portfolios}
 	\label{tab:K_summary_40_appendix}
 	\begin{adjustbox}{max width=\textwidth}
 		\begin{tabular}{l l c c c c}
 			\toprule
 			Portfolio & Setting & Mean & Std & Skew & Kurt \\
 			\midrule
 			Portfolio A & Independent & 5.25 & 2.12 & 0.34 & 3.08 \\
 			& Correlated  & 5.26 & 4.24 & 1.31 & 4.58 \\
 			Portfolio B & Independent & 7.14 & 2.41 & 0.28 & 3.07 \\
 			& Correlated  & 7.15 & 4.85 & 1.12 & 4.02 \\
 			Portfolio C (Diversified) & Independent & 4.44 & 1.98 & 0.38 & 3.06 \\
 			& Correlated  & 4.44 & 3.77 & 1.54 & 5.64 \\
 			Portfolio C -- SaaS & Independent & 4.58 & 2.00 & 0.40 & 3.16 \\
 			& Correlated  & 4.59 & 3.82 & 1.47 & 5.40 \\
 			Portfolio C -- AI & Independent & 4.28 & 1.94 & 0.41 & 3.15 \\
 			& Correlated  & 4.29 & 3.75 & 1.55 & 5.66 \\
 			Portfolio C -- Fintech & Independent & 4.22 & 1.94 & 0.41 & 3.15 \\
 			& Correlated  & 4.24 & 3.64 & 1.59 & 5.98 \\
 			Portfolio C -- Consumer & Independent & 3.92 & 1.88 & 0.44 & 3.16 \\
 			& Correlated  & 3.93 & 3.53 & 1.65 & 6.22 \\
 			Portfolio C -- DevTools & Independent & 4.48 & 1.99 & 0.40 & 3.13 \\
 			& Correlated  & 4.50 & 3.71 & 1.57 & 5.96 \\
 			Portfolio C -- Health & Independent & 3.49 & 1.78 & 0.47 & 3.19 \\
 			& Correlated  & 3.49 & 3.27 & 1.76 & 6.83 \\
 			\bottomrule
 		\end{tabular}
 	\end{adjustbox}
 \end{table}

 \begin{table}[H]
 	\centering
 	\caption{Summary Statistics of the Number of Successful Deals ($K$), 80-Deal Portfolios}
 	\label{tab:K_summary_80_appendix}
 	\begin{adjustbox}{max width=\textwidth}
 		\begin{tabular}{l l c c c c}
 			\toprule
 			Portfolio & Setting & Mean & Std & Skew & Kurt \\
 			\midrule
 			Portfolio A & Independent & 10.83 & 3.02 & 0.24 & 3.05 \\
 			& Correlated  & 10.85 & 8.09 & 1.44 & 5.04 \\
 			Portfolio B & Independent & 14.07 & 3.39 & 0.19 & 2.99 \\
 			& Correlated  & 14.09 & 9.40 & 1.22 & 4.14 \\
 			Portfolio C (Diversified) & Independent & 8.24 & 2.70 & 0.30 & 3.11 \\
 			& Correlated  & 8.21 & 6.92 & 1.77 & 6.48 \\
 			Portfolio C -- SaaS & Independent & 8.84 & 2.79 & 0.28 & 3.05 \\
 			& Correlated  & 8.85 & 7.19 & 1.70 & 6.18 \\
 			Portfolio C -- AI & Independent & 8.58 & 2.75 & 0.29 & 3.08 \\
 			& Correlated  & 8.59 & 7.27 & 1.70 & 6.05 \\
 			Portfolio C -- Fintech & Independent & 8.34 & 2.71 & 0.30 & 3.11 \\
 			& Correlated  & 8.33 & 6.95 & 1.79 & 6.66 \\
 			Portfolio C -- Consumer & Independent & 7.81 & 2.64 & 0.30 & 3.07 \\
 			& Correlated  & 7.82 & 6.77 & 1.81 & 6.71 \\
 			Portfolio C -- DevTools & Independent & 8.58 & 2.74 & 0.29 & 3.08 \\
 			& Correlated  & 8.60 & 7.02 & 1.75 & 6.46 \\
 			Portfolio C -- Health & Independent & 6.97 & 2.51 & 0.34 & 3.11 \\
 			& Correlated  & 6.98 & 6.25 & 1.98 & 7.69 \\
 			\bottomrule
 		\end{tabular}
 	\end{adjustbox}
 \end{table}

 \begin{table}[H]
 	\centering
 	\caption{Portfolio-Level Tail Probabilities, 20-Deal Portfolios}
 	\label{tab:tail_20}
 	\begin{adjustbox}{max width=\textwidth}
 		\begin{tabular}{l l c c c c c}
 			\toprule
 			Portfolio & Setting & $P(K\ge1)$ & $P(K\ge2)$ & $P(K\ge3)$ & $P(K\ge5)$ & $P(K\ge10)$ \\
 			\midrule
 			Portfolio A & Independent & 93.97\% & 75.55\% & 49.20\% & 10.55\% & 0.01\% \\
 			& Correlated  & 80.53\% & 56.49\% & 37.93\% & 18.25\% & 3.20\% \\
 			Portfolio B & Independent & 97.82\% & 88.77\% & 70.49\% & 25.92\% & 0.08\% \\
 			& Correlated  & 89.65\% & 70.71\% & 51.82\% & 27.64\% & 5.90\% \\
 			Portfolio C (Diversified)
 			& Independent & 92.17\% & 71.43\% & 43.71\% & 8.10\% & 0.01\% \\
 			& Correlated  & 78.52\% & 52.99\% & 34.59\% & 16.07\% & 2.53\% \\
 			Portfolio C -- SaaS
 			& Independent & 89.66\% & 64.50\% & 36.01\% & 5.16\% & 0.00\% \\
 			& Correlated  & 74.42\% & 47.45\% & 29.95\% & 14.27\% & 1.66\% \\
 			Portfolio C -- AI
 			& Independent & 88.90\% & 63.24\% & 34.49\% & 4.79\% & 0.00\% \\
 			& Correlated  & 73.32\% & 46.43\% & 28.92\% & 13.61\% & 1.77\% \\
 			Portfolio C -- Fintech
 			& Independent & 87.88\% & 61.00\% & 32.46\% & 4.13\% & 0.00\% \\
 			& Correlated  & 73.42\% & 45.76\% & 27.81\% & 12.39\% & 1.63\% \\
 			Portfolio C -- Consumer
 			& Independent & 86.30\% & 57.61\% & 29.00\% & 3.32\% & 0.00\% \\
 			& Correlated  & 71.09\% & 43.02\% & 25.82\% & 11.42\% & 1.38\% \\
 			Portfolio C -- DevTools
 			& Independent & 88.84\% & 63.32\% & 34.88\% & 4.89\% & 0.00\% \\
 			& Correlated  & 76.54\% & 48.66\% & 29.11\% & 12.31\% & 1.69\% \\
 			Portfolio C -- Health
 			& Independent & 83.17\% & 52.01\% & 23.92\% & 2.34\% & 0.00\% \\
 			& Correlated  & 68.34\% & 39.49\% & 22.58\% & 9.53\% & 1.11\% \\
 			\bottomrule
 		\end{tabular}
 	\end{adjustbox}
 \end{table}
 
 \begin{table}[H]
 	\centering
 	\caption{Portfolio-Level Tail Probabilities, 40-Deal Portfolios}
 	\label{tab:tail_40}
 	\begin{adjustbox}{max width=\textwidth}
 		\begin{tabular}{l l c c c c c}
 			\toprule
 			Portfolio & Setting & $P(K\ge1)$ & $P(K\ge2)$ & $P(K\ge3)$ & $P(K\ge10)$ & $P(K\ge20)$ \\
 			\midrule
 			Portfolio A & Independent & 99.63\% & 97.61\% & 91.28\% & 2.97\% & 0.00\% \\
 			& Correlated  & 95.22\% & 84.54\% & 70.97\% & 15.34\% & 0.64\% \\
 			Portfolio B & Independent & 99.98\% & 99.64\% & 98.23\% & 16.18\% & 0.00\% \\
 			& Correlated  & 98.61\% & 94.23\% & 86.47\% & 25.10\% & 2.38\% \\
 			Portfolio C (Diversified) & Independent & 99.08\% & 94.63\% & 83.63\% & 1.04\% & 0.00\% \\
 			& Correlated  & 93.52\% & 79.94\% & 64.05\% & 10.25\% & 0.32\% \\
 			Portfolio C -- SaaS & Independent & 99.27\% & 95.46\% & 85.41\% & 1.34\% & 0.00\% \\
 			& Correlated  & 94.12\% & 81.35\% & 65.41\% & 11.37\% & 0.37\% \\
 			Portfolio C -- AI & Independent & 98.92\% & 93.93\% & 81.87\% & 0.81\% & 0.00\% \\
 			& Correlated  & 92.52\% & 78.03\% & 61.29\% & 10.08\% & 0.31\% \\
 			Portfolio C -- Fintech & Independent & 98.82\% & 93.50\% & 80.91\% & 0.75\% & 0.00\% \\
 			& Correlated  & 92.87\% & 78.58\% & 61.88\% & 9.29\% & 0.30\% \\
 			Portfolio C -- Consumer & Independent & 98.36\% & 91.52\% & 76.86\% & 0.48\% & 0.00\% \\
 			& Correlated  & 91.07\% & 74.83\% & 57.53\% & 8.12\% & 0.23\% \\
 			Portfolio C -- DevTools & Independent & 99.14\% & 94.96\% & 84.10\% & 1.12\% & 0.00\% \\
 			& Correlated  & 94.29\% & 81.86\% & 65.85\% & 9.89\% & 0.41\% \\
 			Portfolio C -- Health & Independent & 97.47\% & 87.63\% & 68.93\% & 0.20\% & 0.00\% \\
 			& Correlated  & 88.95\% & 70.36\% & 51.78\% & 6.32\% & 0.13\% \\
 			\bottomrule
 		\end{tabular}
 	\end{adjustbox}
 \end{table}
 
 \begin{table}[H]
 	\centering
 	\caption{Portfolio-Level Tail Probabilities, 80-Deal Portfolios}
 	\label{tab:tail_80}
 	\begin{adjustbox}{max width=\textwidth}
 		\begin{tabular}{l l c c c c c}
 			\toprule
 			Portfolio & Setting & $P(K\ge1)$ & $P(K\ge2)$ & $P(K\ge3)$ & $P(K\ge20)$ & $P(K\ge40)$ \\
 			\midrule
 			Portfolio A & Independent & 100.00\% & 99.99\% & 99.92\% & 0.39\% & 0.00\% \\
 			& Correlated  & 99.45\% & 97.52\% & 93.52\% & 14.13\% & 0.59\% \\
 			Portfolio B & Independent & 100.00\% & 100.00\% & 99.998\% & 5.96\% & 0.00\% \\
 			& Correlated  & 99.88\% & 99.28\% & 97.72\% & 22.44\% & 2.01\% \\
 			Portfolio C (Diversified) & Independent & 99.98\% & 99.83\% & 99.20\% & 0.01\% & 0.00\% \\
 			& Correlated  & 98.39\% & 93.61\% & 85.74\% & 8.08\% & 0.16\% \\
 			Portfolio C -- SaaS & Independent & 99.99\% & 99.91\% & 99.51\% & 0.03\% & 0.00\% \\
 			& Correlated  & 98.86\% & 95.08\% & 88.44\% & 9.22\% & 0.26\% \\
 			Portfolio C -- AI & Independent & 99.99\% & 99.90\% & 99.40\% & 0.02\% & 0.00\% \\
 			& Correlated  & 98.42\% & 93.66\% & 86.17\% & 9.43\% & 0.23\% \\
 			Portfolio C -- Fintech & Independent & 99.99\% & 99.83\% & 99.23\% & 0.02\% & 0.00\% \\
 			& Correlated  & 98.35\% & 93.80\% & 86.32\% & 8.11\% & 0.21\% \\
 			Portfolio C -- Consumer & Independent & 99.99\% & 99.75\% & 98.73\% & 0.01\% & 0.00\% \\
 			& Correlated  & 97.84\% & 92.25\% & 83.41\% & 7.62\% & 0.16\% \\
 			Portfolio C -- DevTools & Independent & 99.99\% & 99.88\% & 99.39\% & 0.03\% & 0.00\% \\
 			& Correlated  & 98.79\% & 94.89\% & 87.99\% & 8.64\% & 0.22\% \\
 			Portfolio C -- Health & Independent & 99.95\% & 99.45\% & 97.49\% & 0.00\% & 0.00\% \\
 			& Correlated  & 97.12\% & 90.03\% & 79.53\% & 5.77\% & 0.08\% \\
 			\bottomrule
 		\end{tabular}
 	\end{adjustbox}
 \end{table}

\end{document}